\documentclass[prd,showpacs,superscriptaddress]{revtex4-1}
\usepackage{epsfig}
\usepackage{mathrsfs}
\usepackage{amsmath}
\usepackage{amssymb}
\usepackage{color}
\usepackage{graphicx}
\usepackage{dcolumn}
\usepackage{bm}

\begin{document}

\title{Interpretation of the measurements \\
of total, elastic and diffractive cross sections at LHC}

\author{Paolo Lipari}
\email{paolo.lipari@roma1.infn.it}
\affiliation{INFN sezione 
Roma ``La Sapienza''} 

\author{Maurizio Lusignoli}
\email{maurizio.lusignoli@roma1.infn.it}
\affiliation{Dipartimento di Fisica and sezione INFN,
Universit\`a di Roma ``La Sapienza''}


\begin{abstract}
Recently at LHC one has obtained measurements of the total, elastic 
and diffractive cross sections in $pp$ collisions at very high energy. 
The total cross section is in good agreement with predictions based on 
a leading behavior $\sigma_{\rm tot} (s) \propto (\ln s/s_0)^2$,
on the other hand the elastic cross section is 
lower than most expectations and the diffractive cross section is higher.
It is remarkable that the ratio 
 $(\sigma_{\rm el} + \sigma_{\rm diff})/\sigma_{\rm tot}$ calculated combining the
results of the TOTEM and ALICE detectors is $0.495^{+0.05}_{-0.06}$,
very close to the maximum theoretically allowed value of 1/2 known
as the Miettinen Pumplin bound.
In this work we discuss these results using the frameworks of
single and multi--channel eikonal models, and outline the main
difficulties for a consistent interpretation of the data.
\end{abstract}


\maketitle

\section{Introduction}
\label{sec:intro}
The data obtained at the CERN LHC has allowed to significantly 
extend the energy range for the study of proton--proton collisions. 
The main focus of the LHC program is the study of 
rare processes and the search for new physics phenomena,
but they also allow to measure 
``average properties'' of the $pp$ interactions, such as the 
total and elastic cross section, multiplicity distributions 
and inclusive spectra in different kinematical variables
(energy, rapidity, pseudo--rapidity or transverse momentum).
These studies have an intrinsic interest because they test 
our understanding of the strong interactions, and at the same time 
they are important for the interpretation of the observations
of ultra high energy cosmic rays (UHECR).
The spectrum of these particles extends to 
$E_{\rm lab} \simeq 10^{20}$~eV, that corresponds to a nucleon--nucleon
c.m. energy $\sqrt{s} \simeq 433$~TeV, and 
the modeling of their atmospheric showers 
requires an extrapolation of the LHC results that should be 
based on a reasonably robust theoretical understanding of the 
underlying hadronic physics.

The TOTEM detector
\cite{Antchev:2011vs,totem-a,Antchev:2011zz,totem-b,totem-c,totem-d}
 has obtained measurements of the $pp$ total and
elastic cross sections at $\sqrt{s} = 7$ and 8~TeV. These results
extend considerably the energy range where these cross sections 
are known, and constitute a very important constraint 
for the extrapolations of 
$\sigma_{\rm tot}^{pp}$
and $\sigma_{\rm el}^{pp}$ up to the UHECR energy range.
Measurements of the inelastic cross sections at 
$\sqrt{s} = 7$~TeV have also
been obtained by the ALICE, ATLAS and CMS detectors
\cite{alice-sigma,atlas-sigma,cms-sigma}.
It should be noted that the interaction length of high energy protons
and nuclei in air is determined by the combination of the total
and elastic cross sections, so that an understanding of the energy dependence
of both quantities is needed.

The ALICE collaboration has also published measurements of the 
inelastic diffraction cross section at $\sqrt{s} = 0.9$, 2.3 and 7~TeV
in the form of the ratios 
$\sigma_{SD}/\sigma_{\rm inel}$ and
$\sigma_{DD}/\sigma_{\rm inel}$, where SD (DD) stands for single 
(double) diffraction. In inelastic diffraction one
(or both) of the colliding protons is excited
into a state with the same internal quantum numbers of the initial particle. 
Summing over single and double diffraction, the diffractive 
processes account for
approximately one third of the inelastic cross section.
It is interesting to note that the 
sum of the elastic and diffraction cross
sections is consistent with the saturation of the theoretical bound 
established by Miettinen and Pumplin 
\cite{Miettinen:1978jb}, according to which 
$\sigma_{\rm diff} + \sigma_{\rm el} \le\sigma_{\rm tot}/2$. 
It is also intriguing that at LHC energies 
the elastic and diffractive cross section are approximately equal.

A good knowledge of the size of the diffraction 
cross section is phenomenologically important 
for the modeling of UHECR showers because the properties
of the final state in diffractive events are significantly different 
than for the rest of inelastic interactions.
From the theoretical point of  view a good description
of diffraction is an important test.

The observations of ALICE, ATLAS, CMS and LHCb have also 
provided many observations of the properties of particle production 
in inclusive (or ``minimum bias'') $pp$ interactions. 
These observations show that the average charged multiplicity,
and the average density of particles per unit
of pseudorapidity in the central region increases with energy
rapidly (faster than most predictions), and have large fluctuations 
(broader than most predictions). 
These measurements are confined to the central region
of phase space, and do not include particles emitted 
at very small angles with respect to the beam directions.
Very valuable observations of the energy spectra
of photons and neutral pions emitted at
very small angles have been obtained by the LHCf detector \cite{Adriani:2011nf}.

In our view, all properties of high energy hadronic interactions:
total, elastic and diffractive cross sections,
average multiplicities, inclusive energy spectra are
associated to the same fundamental underlying dynamics,
and require a comprehensive and consistent description.
The fundamental idea underlying such a comprehensive 
description of the properties of inclusive hadronic interactions,
is that hadrons are composite objects
formed by an ensemble of quarks and gluons, and that in a single
$pp$ (or more in general hadron--hadron) 
collision there are several parton--parton interactions.
With increasing energy a larger number of parton--parton
collisions becomes possible, and this drives at the same time
the growth of the hadron--hadron cross sections and the growth 
of the particles multiplicity in the final state
(with a corresponding softening of the inclusive energy spectra).
A comprehensive model should therefore be able to relate the 
energy dependence of the parton--parton 
and hadron--hadron cross sections, and then also
connect the underlying parton structure of the collision to the 
observable final state particles.
The LHC data are of course essential for the development
of such a comprehensive model for hadronic interactions
at high energy.

The LHC  results
on the total, elastic and diffractive cross sections 
have been the object of several theoretical  studies,
see for example \cite{Ryskin:2012az,
Gotsman:2012rq,Gotsman:2012rm,Grau:2012wy,Kopeliovich:2012yy,Fagundes:2012rr}.

This work is organized as follows: in the next section 
we list and briefly discuss the measurements of the 
total, elastic and diffractive cross sections at LHC. 
In section 3 we introduce the definitions of the opacity 
and eikonal functions $\Gamma(b,s)$ and $\chi(b,s)$, and 
estimate these functions from the data on the elastic
cross sections  obtained by TOTEM at $\sqrt{s} = 7$~TeV, and
compare with data at lower energy.
In section 4 and 5 we try to interpret 
these results in the frameworks
of single and  multi--channel eikonal models.
The last section contains some discussion.

\section{Cross section measurements at LHC}
\label{sec:LHC}
Recently the TOTEM collaboration has measured 
\cite{Antchev:2011vs,totem-a,Antchev:2011zz,totem-b,totem-c,totem-d}
the total, elastic and inelastic cross sections in
$pp$ collisions at $\sqrt{s} = 7$ and 8~TeV. 
The cross sections at $\sqrt{s} = 7$~TeV have been obtained using
two different methods. The first one 
\cite{Antchev:2011vs,totem-a,Antchev:2011zz} makes use 
of the luminosity measurements of the CMS detector, 
and is based on the measurement of elastic scattering, with the 
total cross section obtained via the optical theorem;
the second one \cite{totem-b,totem-c,totem-d} is independent
from a luminosity measurements and relies on measurements of
elastic an inelastic events (and again the optical theorem).
At $\sqrt{s} = 8$~TeV the cross sections measurements 
\cite{totem-d} have been presented
only for the luminosity independent method.

At $\sqrt{s} = 7$~TeV, the TOTEM collaboration has measured 
the differential elastic scattering cross section 
 in the range $0.005 \le |t| \le 0.47$~GeV$^2$,
with $t = (p - p^\prime)^2$ the squared momentum transfer
 \cite{Antchev:2011zz,Antchev:2011vs,totem-a}.
Extrapolating the data to $t = 0$, and using the CMS luminosity 
measurement, the collaboration obtains 
\cite{Antchev:2011vs} a total elastic cross section 
$\sigma_{\rm el} = 24.8 \pm 1.2$ (or $25.4 \pm 1.1$ in \cite{totem-a}).
The data on the elastic cross section 
are shown in fig.~\ref{fig:totem-el}.
For a comparison to lower energy data,
 fig.~\ref{fig:isr-el} shows the data 
\cite{Nagy:1978iw,Ambrosio:1982zj,Breakstone:1984te}
on elastic scattering obtained at ISR for $\sqrt{s} = 53$~GeV,
where the data extends to a very broad range in $t$.

The differential elastic cross section
can be written in terms of the scattering amplitude $F_{\rm el} (t,s)$ as:
\begin{equation}
\frac{d \sigma_{\rm el}}{dt} (t, s) = 
\pi \; \left | F_{\rm el} (t, s) \right |^2 \;.
\label{eq:gen_dsig}
\end{equation}
The optical theorem gives a relation between the total cross section 
and the imaginary part of the forward elastic scattering amplitude:
\begin{equation}
\sigma_{\rm tot}(s) = 
4 \, \pi \; \Im [F_{\rm el} (0, s) ] = 
\frac{4 \, \sqrt{\pi}}{\sqrt{1 + \rho^2}} 
 \;
\left [\left . \frac{d\sigma_{\rm el}}{dt} \right |_{t = 0} \right ]^{1/2} 
\label{eq:optical}
\end{equation}
where we have used the notation 
$F_{\rm el} (0, s) = \Im \left[F_{\rm el} (0, s) \right ] \; (i + \rho)$
so that $\rho$ is the ratio of the real to imaginary part
of the forward elastic scattering amplitude.

Extrapolating the elastic cross section measurements to $t =0$
and using the prediction 
 $\rho \simeq 0.145 \pm 0.007$ at $\sqrt{s} = 7$~TeV
from the COMPETE group \cite{Cudell:2002xe},
the TOTEM collaboration arrives to the result \cite{Antchev:2011vs} 
$\sigma_{\rm tot} = 98.3 \pm 2.8$~mb
(or $\sigma_{\rm tot} = 98.6 \pm 2.2$~mb in \cite{totem-a}).
By subtraction one obtains
$\sigma_{\rm inel} = 73.5 \pm 1.6$~mb (or $73.2 \pm 1.3$~mb).

The luminosity independent method relies on the 
measurement of integrated number of elastic and inelastic events 
($N_{\rm el}$ and $N_{\rm inel}$) combined with 
the measurement of the quantity $\left . dN_{\rm el}/dt \right |_{t =0}$.
Using this method at $\sqrt{s} = 7$~TeV \cite{totem-b,totem-c} 
the TOTEM collaboration obtains at 7~TeV:
$\sigma_{\rm tot} = 98.0 \pm 2.5$,
$\sigma_{\rm el} = 25.1 \pm 1.1$ and
$\sigma_{\rm inel} = 72.9 \pm 1.5$, in good agreement with the 
results obtained by the previous method.
The results can be combined with the luminosity measurement
of CMS to estimate $\rho = 0.145 \pm 0.091$ in excellent agreement
with the COMPETE prediction.

At $\sqrt{s} = 8$~TeV \cite{totem-d} the luminosity independent method gives
$\sigma_{\rm tot} = 101.7 \pm 2.9$,
$\sigma_{\rm el} = 27.1 \pm 1.4$ and
$\sigma_{\rm inel} = 74.7 \pm 1.7$. 

For an understanding of the dynamics that determines the
total and elastic cross section it is important to also
measure the inelastic diffraction cross section.
The ALICE collaboration 
\cite{alice-sigma} has 
recently published measurements of this
component of the $pp$ cross section at LHC. 
The rate of diffractive events 
is estimated from a study of gaps in the pseudorapidity distributions
of charged particles. 
For a diffractive mass $M_X^2 < 200$ GeV$^2$ the fraction of single 
diffraction processes in inelastic collisions 
 at $\sqrt{s} = 0.9$, 2.36 and 7~TeV is estimated as:
$\sigma_{\rm SD}/\sigma_{\rm inel} = 0.21 \pm 0.03$, 
$0.20^{+0.07}_{-0.08}$, 
and $0.20^{+0.04}_{-0.07}$.
For double diffraction processes 
(with a pseudorapidity gap $\Delta\eta > 3$),
for the same c.m. energies,
ALICE finds: 
$\sigma_{\rm DD}/\sigma_{\rm inel} = 0.11 \pm 0.03$, 
$0.12 \pm 0.05$, 
and $0.12^{+0.05}_{-0.04}$.

Support for a large cross section for inelastic diffraction
also comes from a comparison of the measurements of
the inelastic cross sections at $\sqrt{s} = 7$~TeV 
obtained by the ATLAS \cite{atlas-sigma} and
CMS \cite{cms-sigma} collaborations, with the 
TOTEM results.
The measurements of $\sigma_{\rm inel}$ of ATLAS and CMS are 
based on a direct method, and require
a model dependent correction to take into account the fact 
that an important fraction of the diffractive events is not observable.
On the other hand the measurement of 
the inelastic cross section in TOTEM
that is obtained as the difference between 
$\sigma_{\rm tot}$ and $\sigma_{\rm el}$ is
independent from the size of the diffractive cross section.
To reconcile the estimates of $\sigma_{\rm inel}$ 
of all experiments it is necessary to include 
a large $\sigma_{\rm diff}$.

Combining the results of ALICE and TOTEM one can 
estimate at $\sqrt{s} = 7$~TeV the ratios
\begin{equation}
\frac{\sigma_{\rm diff}}{\sigma_{\rm tot}} \simeq 0.24 ~^{+0.05}_{-0.06}
\end{equation}
and
\begin{equation}
\frac{\sigma_{\rm el} + \sigma_{\rm diff}}{\sigma_{\rm tot}} 
= 0.495^{+0.05}_{-0.06}~.
\label{eq:rsigexp}
\end{equation}
Miettinen and Pumplin \cite{Miettinen:1978jb} have argued that the sum
of the diffractive and elastic cross section 
must respect the bound:
\begin{equation}
\frac{\sigma_{\rm el} + \sigma_{\rm diff}}{\sigma_{\rm tot}} 
\le \frac{1}{2} \;.
\label{eq:pumplin-bound1}
\end{equation}
A derivation of the bound is also given below
 in section~\ref{sec:multichannel}.
The combined measurements of TOTEM and ALICE
indicate that the Miettinen--Pumplin
bound is close to saturation at LHC energies.
This is a remarkable result that had not been predicted, 
and is in need of a convincing dynamical explanation.


It is also interesting that the elastic and diffractive cross
sections at $\sqrt{s} = 7$~TeV are consistent with being equal:
\begin{equation}
\frac{\sigma_{\rm diff}}{\sigma_{\rm el}} = 0.952^{+0.20}_{-0.24}~.
\end{equation}
This is likely to be only an accident, but it is possible 
to speculate whether, with increasing $\sqrt{s}$,
the ratio $\sigma_{\rm diff}/\sigma_{\rm el}$ will stabilize to
a constant value or decrease.

\section{Opacity and eikonal functions}
\label{sec:opacity}
In order to construct models of the total and elastic
cross sections based on the partonic structure of the colliding
hadrons it is useful to study the collisions in impact parameter space.
This allows the definition and construction of 
quantities (such as the the opacity and eikonal functions)
that can then be more directly interpreted in terms of elementary parton
interactions.

The elastic scattering amplitude $F_{\rm el} (t,s)$
can be written as a two--dimensional integral over impact parameter:
\begin{equation}
F_{\rm el} (t,s) 
 = i\; \int \frac{d^2b}{2 \, \pi} ~ e^{i \vec{q}.\vec{b}}~
\Gamma (b, s) \; .
\label{eq:gen_ampl}
\end{equation}
In this equation the 2--dimensional vector
$\vec{q}$ gives the spatial components of the transverse
momentum.
At high energy and small $|t|$, to a very good approximation one has:
$|\vec{q}|^2 = -t$.
The quantity $\Gamma (b,s)$ is the opacity function
that can be written 
in terms of the eikonal function $\chi(b,s)$:
\begin{equation}
\Gamma (b,s) = 1- e^{-\chi(b,s)} \;.
\label{eq:eik_def}
\end{equation}
Substituting the expression (\ref{eq:gen_ampl}) for the amplitude
into (\ref{eq:gen_dsig}) and integrating over all $t$ (over $d^2q$) 
one obtains:
\begin{equation}
\sigma_{\rm el}(s) = 
 \int d^2b ~\left | \Gamma (b,s) \right |^2 ~.
\label{eq:siggen_el}
\end{equation}
From the optical theorem one has:
\begin{equation}
\sigma_{\rm tot}(s) = 
 2 \; \int d^2b ~\Re \left [\Gamma (b,s) \right ] ~.
\label{eq:siggen_tot}
\end{equation}
Combining equations 
(\ref{eq:siggen_el}) and (\ref{eq:siggen_tot}) 
one also obtains:
\begin{equation}
\sigma_{\rm inel}(s) = \int d^2b ~\left \{1 - \left |1 - \Gamma (b,s) \right |^2
 \right \} ~.
\label{eq:siggen_inel}
\end{equation}

In the following we will make the approximation to neglect
the real part of the elastic scattering amplitude.
This is incorrect because in this situation the 
amplitude has not the required analiticity properties.
The approximation is reasonable, since 
$\rho(s)$, the ratio of the real to imaginary part of 
the amplitude at $t=0$ is small.
The assumption of a purely imaginary amplitude
implies that the opacity and eikonal functions are real,
and a physical interpretation of these objects in terms
of parton--parton interactions becomes easier.

The width of $\Gamma(b,s)$, 
or more precisely the average value $\langle b^2 (s) \rangle$,
is directly proportional to the slope $B_{\rm el}(s)$ of the differential
cross section at $t= 0$ ($d\sigma_{\rm el}/dt \propto e^{B_{\rm el} \,t}$). 
From the definition: 
\begin{equation}
B_{\rm el}(s) = \left [ \left (\frac{d\sigma_{\rm el}}{dt} ~\right )^{-1}
\frac{d}{dt} \left ( \frac{d \sigma_{\rm el}}{dt} \right ) \right ]_{t=0} 
~,
\end{equation}
and using equations (\ref{eq:gen_dsig}) and (\ref{eq:gen_ampl}) 
one can derive the expression:
\begin{equation}
B_{\rm el}(s) = 
\left \{ \int d^2b ~ \frac{b^2}{2} ~ \Gamma (b,s) 
\right \}
\times 
\left \{ \int d^2b ~ 
\Gamma (b,s) \right \}^{-1} = 
\frac{\langle b^2 \rangle } {2} \;,
\label{eq:Bdef}
\end{equation}

\begin{figure} [hbt]
\includegraphics[width=14.0cm]{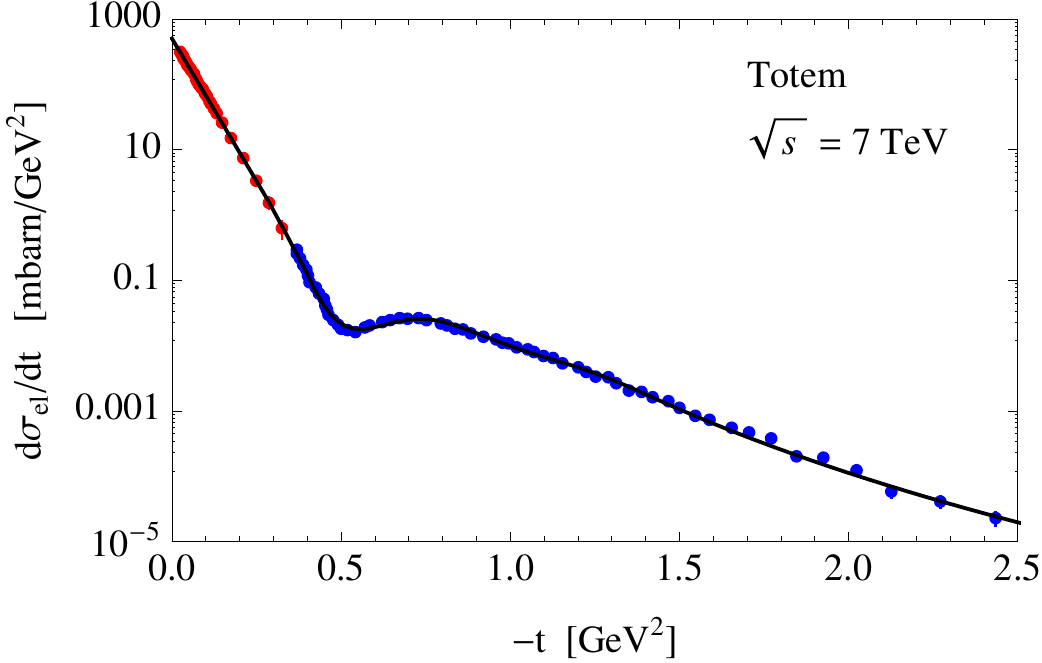}

\caption {\footnotesize
Measurements of the elastic cross section 
in $pp$ collisions at $\sqrt{s} = 7$~TeV
obtained by the TOTEM collaboration \protect\cite{Antchev:2011vs,totem-a,Antchev:2011zz}.
The line is a fit to the data.
\label{fig:totem-el} }
\end{figure}

\begin{figure} [hbt]
\includegraphics[width=14.0cm]{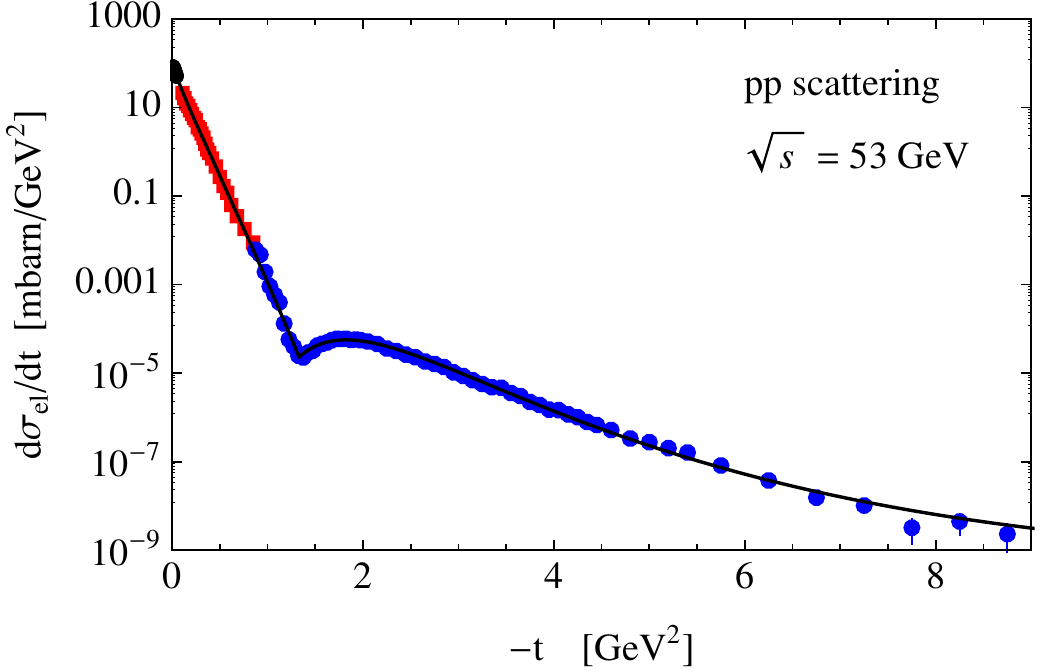}

\caption {\footnotesize
Measurements of the elastic cross section 
in $pp$ collisions at $\sqrt{s} = 53$~GeV 
obtained at ISR  \protect\cite{Nagy:1978iw,Ambrosio:1982zj,Breakstone:1984te}.
The line is a fit to the data.
\label{fig:isr-el} }
\end{figure}

The elastic scattering cross section $d\sigma_{\rm el}/dt$ 
is well described (see fig.~\ref{fig:totem-el} and~\ref{fig:isr-el})
by a simple exponential of constant slope for $|t|$ not too
large. Using the approximation that the 
form $d\sigma_{\rm el}/dt \propto e^{B_{\rm el} \, t}$ is valid for all 
$t$ one can write the elastic cross section as:
\begin{equation}
\sigma_{\rm el} (s) \simeq \pi ~ \frac{\left | F_{\rm el} (0,s) \right |^2}
{B_{\rm el}(s)}
= \frac{\sigma_{\rm tot}^2 \;
(1 + \rho^2) }{16 \, \pi \, B_{\rm el}} ~,
\label{eq:optical1}
\end{equation}
(in the second equality we have used the optical theorem).
This equation relates 
$\sigma_{\rm tot}$, $\sigma_{\rm el}$ and the slope $B_{\rm el}$
(with a smaller role played by $\rho$).

The opacity function $\Gamma (b,s)$ can be obtained 
from the elastic scattering amplitude inverting 
equation (\ref{eq:gen_ampl}):
\begin{equation}
\Gamma (b,s) = -i \; \int_0^\infty d|t| ~J_0 (b\,\sqrt{|t|})~F_{\rm el} (t,s)
\label{eq:gamma_comp}
\end{equation}
In general it is not possible to use the above equation 
to evaluate numerically $\Gamma(b,s)$ from a measurement of the differential
elastic scattering cross section 
because one lacks information about the phase of amplitude.
However, it is known that the elastic scattering
amplitude is in first approximation purely imaginary,
and also that at $t=0$ this imaginary part is positive,
since it is related by the optical theorem
to the total cross section.
If one makes the approximation to neglect the real
part of the amplitude and in addition
one assumes that the derivative 
(with respect to $t$) of the amplitude has no discontinuities,
the phase ambiguity is resolved.
The (purely real) opacity function can then be 
obtained with the numerical integration:
\begin{equation}
\Gamma (b,s) = \frac{1}{\sqrt{\pi}} \; 
\int_0^\infty d|t| ~J_0 (b\, \sqrt{|t|})~
\sqrt{\frac{d\sigma_{\rm el}}{dt}}\; {\rm sign}(t,s) \;.
\label{eq:gamma_comp1}
\end{equation}
(where the sign is $+1$ at $t=0$, and changes at each zero 
of the cross section).

\begin{figure} [hbt]
\includegraphics[width=14.0cm]{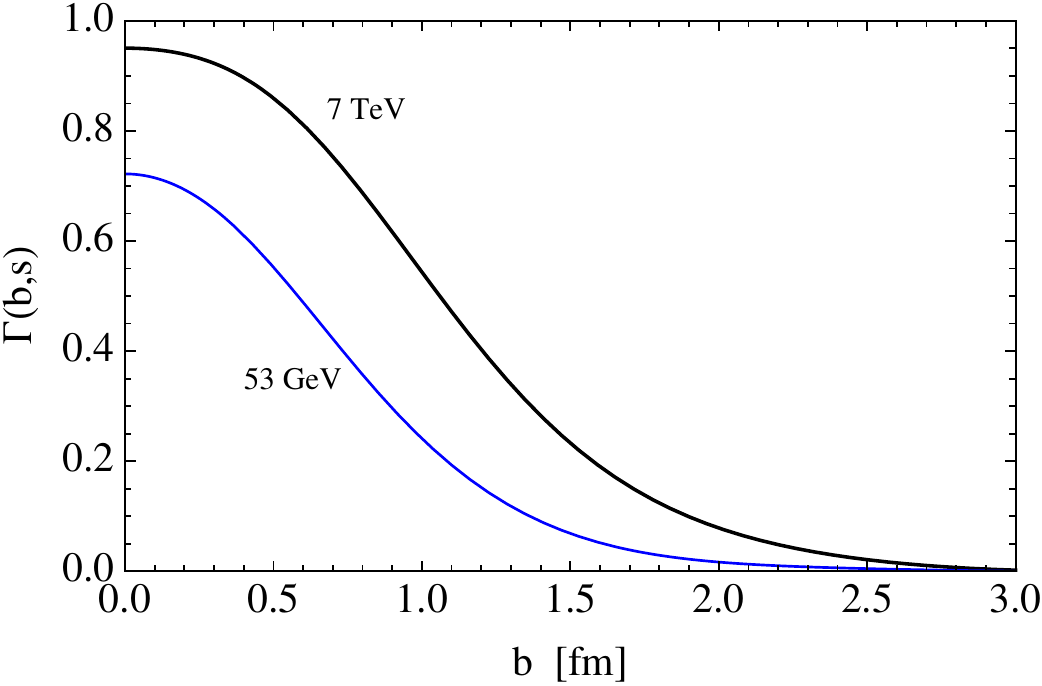}
\caption {\footnotesize
Opacity functions $\Gamma(b,s)$ 
calculated from the measurements of the 
differential elastic cross sections in 
$pp$ collisions at $\sqrt{s}= 53$ and 7000~GeV.
The opacity functions are obtained from the fits to the data
shown fig.~\ref{fig:totem-el} and~\ref{fig:isr-el}
using equation (\ref{eq:gamma_comp1}).
The corresponding eikonal functions are shown in fig.~\ref{fig:chi_exp}.
\label{fig:gamma_exp} }
\end{figure}

We have used equation (\ref{eq:gamma_comp1})
to estimate numerically the profile and
eikonal function from the experimental data.
Using for $d\sigma_{\rm el}/dt$ the fit to the data shown in
fig.~\ref{fig:totem-el} and~\ref{fig:isr-el} one obtains
the profile functions $\Gamma (b,s)$ shown in fig.~\ref{fig:gamma_exp}.
The data covers only a finite range of $t$, but the extrapolation
to large $-t$ introduces a negligible error when $b$ is not too large.
With increasing $s$ the opacity function becomes larger
and broader, remaining (as expected) 
always in the range $0 \le \Gamma (b,s) \le 1$.

\begin{figure} [hbt]
\includegraphics[width=14.0cm]{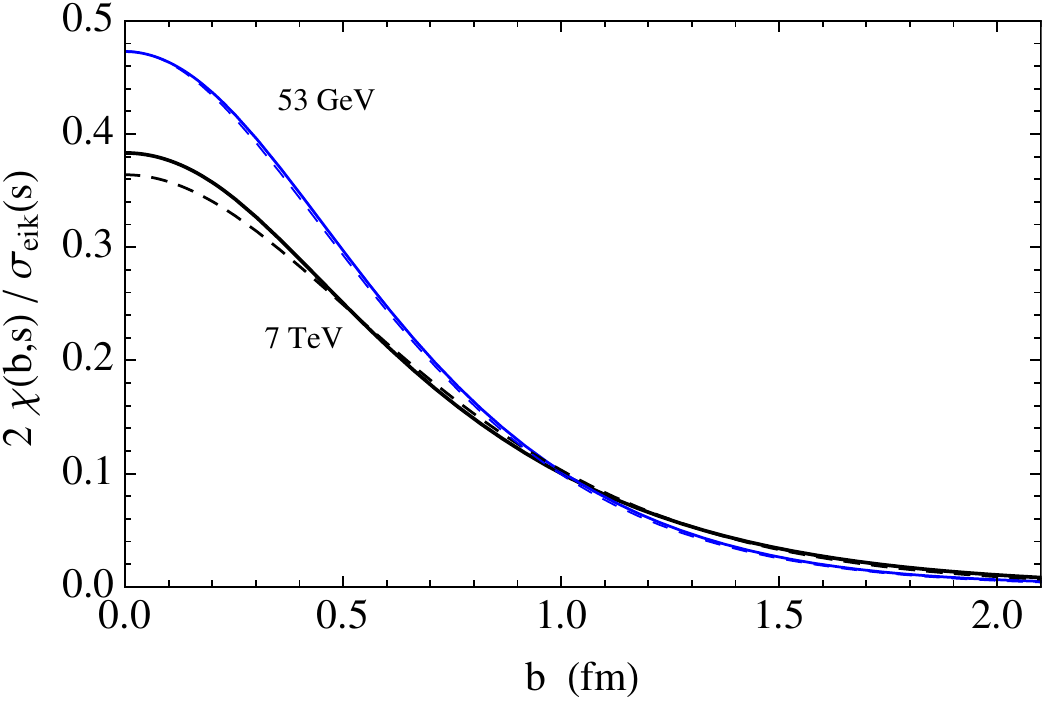}
\caption {\footnotesize
Shape of the eikonal  function $A_\chi(b,s) = 2 \, \chi(b,s)/\sigma_{\rm eik}(s)$
at $\sqrt{s} = 53$~GeV and 7~TeV.
The eikonal function was obtained  from the fits to the data 
on  elastic scattering  shown  in fig.~\ref{fig:totem-el}
and~\ref{fig:isr-el}.
The corresponding opacity functions are shown in fig.~\ref{fig:gamma_exp}.
\label{fig:chi_exp} }
\end{figure}

The corresponding eikonal functions can be calculated from the definition
(\ref{eq:eik_def}) and are shown in fig.~\ref{fig:chi_exp}.
Without loss of generality the function $\chi(b,s)$
 can be written as the product:
\begin{equation}
\chi(b,s) = \frac{1}{2} \; A_\chi (b,s) \; \sigma_{\rm eik} (s)
\label{eq:chi_decomp}
\end{equation}
where the quantity $\sigma_{\rm eik}(s)$ (with the dimensions 
of a cross section) can be obtained 
performing the integration:
\begin{equation}
\sigma_{\rm eik} (s) = 2 \; \int d^2 b~ \chi(b,s) \;,
\label{eq:sig_eik}
\end{equation}
 and the function $A_\chi(b,s)$ (with the dimensions of the inverse
of a cross section) is normalized to unity:
\begin{equation}
\int d^2 b~ A_\chi (b,s) = 1 \;.
\end{equation}

For the opacity functions shown in figures~\ref{fig:gamma_exp} 
and~\ref{fig:chi_exp} one obtains $\sigma_{\rm eik} (s) \simeq 55 \pm 1$~mb 
at $\sqrt{s} = 53$~GeV
and $\sigma_{\rm eik} (s) \simeq 156\pm 14$~mb at $\sqrt{s} = 7$~TeV.
The shapes of the eikonal functions are shown in fig.~\ref{fig:chi_exp}.
With increasing $\sqrt{s}$ the function $A_\chi(b,s)$ becomes
broader: At $\sqrt{s} = 53$~GeV 
one has $\sqrt{\langle b^2 \rangle} \simeq 0.93 \pm 0.01$~fm, 
 at 7~TeV $\sqrt{\langle b^2 \rangle} \simeq 1.09\pm 0.02$~fm.

To have an analytic approximation for the 
$b$ dependence of the eikonal function we have used 
the form:
\begin{equation}
A_\chi(b,s) \simeq A_*[b,r_0] = \frac{1}{96 \, \pi \; r_0^2} \; 
\left [ \frac{b}{r_0} \right ]^3 \; K_3 \left ( \frac{ b}{r_0} \right )
\label{eq:App}
\end{equation}
that depends on the single parameter $r_0$, that must be considered
as a function of $s$, and is proportional to the width
of the distribution:
\begin{equation}
\left \langle b^2 (s) \right \rangle = 16 \, r_0^2(s) ~.
\end{equation}
The expression (\ref{eq:App}) 
corresponds to the geometrical overlap of two normalized, spherical 
exponential distributions ($\rho (r) \propto e^{-r/r_0}$) separated by the 
transverse distance $b$:
\begin{equation}
A_*(b) = 
\int d^3 r_1 \; \int d^3 r_2 ~
\rho(r_1)\; \rho (r_2)
~\delta^{(2)} (\vec{r}_{\perp 1} - \vec{r}_{\perp 2} + \vec{b}) ~.
\end{equation}
This can be verified noting that the Fourier transform
of an exponential distribution is $(1+ q^2 \, r_0^2)^{-2}$, and 
the Fourier transform of expression (\ref{eq:App}) is: 
\begin{equation}
\hat{A}_*(q) = (1 + q^2 \, r_0^2)^{-4} \;.
\end{equation}
The proton electromagnetic form factor has also
the form $(1+ R_p^2 \, q^2)^{-2}$ (with $R_p \simeq 0.234$~fm),
and on this basis the functional form (\ref{eq:App}),
(with the parameter $r_0$ fixed at the value $r_0 = R_p$)
was proposed by Durand and Pi \cite{Durand:1988cr} as the $b$ dependence
of the eikonal function, and then used in several other works,
sometimes leaving $r_0$ as a free parameter.

In fig.~\ref{fig:chi_exp} the dashed lines show the 
function $A_* (b, r_0)$ with the parameter $r_0$ chosen to
reproduce the width of the eikonal function $\chi(b,s)$
obtained numerically.
The functional form (\ref{eq:App}) is an excellent description of the 
shape of the eikonal at $\sqrt{s} = 53$~GeV and 
remains a reasonably good description 
at $\sqrt{s} = 7$~TeV, where some small deviations become apparent.

\begin{figure} [hbt]
\includegraphics[width=14.0cm]{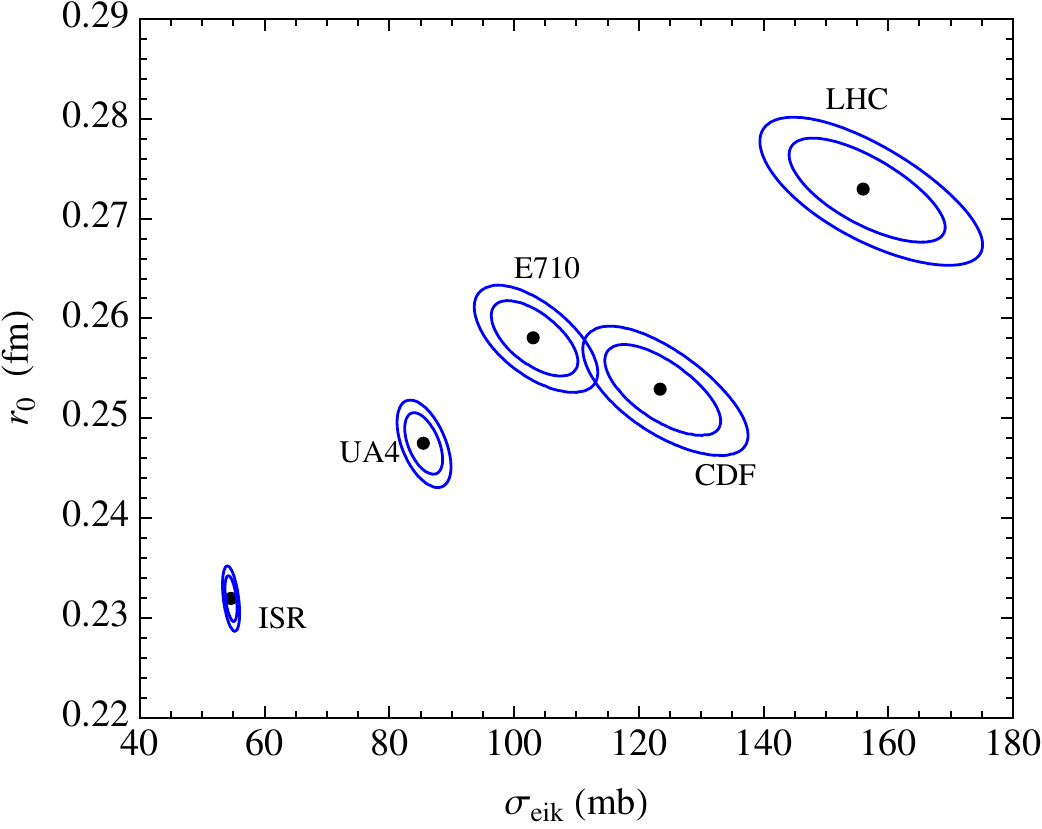}
\caption {\footnotesize 
Regions in the plane $\{\sigma_{\rm eik},r_0\}$ that 
corresponds to the measurements of
$\sigma_{\rm tot}$ and $B_{\rm el}$ for
$pp$ and $\overline{p}p$ collisions at different
c.m. energies.
The different regions are
for $pp$ collisions at 
$\sqrt{s} = 53$~GeV (ISR),
and 
$\sqrt{s} = 7$~TeV (LHC);
and $\overline{p}p$ collisions at
$\sqrt{s} = 546$~GeV (UA4),
and $\sqrt{s} = 1.8$~TeV (CDF) and (E710).
\label{fig:error} }
\end{figure}

Allowed regions 
(at 68\% and 90\% C.L.) for the parameters $\sigma_{\rm eik}$ and $r_0(s)$
that describe the experimental data are shown in fig.~\ref{fig:error}.
In the figure we also show the results for $\overline{p}p$ scattering
at $\sqrt{s} = 0.546$ and 1.8~TeV.
At $\sqrt{s} = 1.8$~TeV the measurements of CDF 
\cite{Abe:1993xx,Abe:1993xy} and E811 \cite{Avila:1998ej}
are in poor agreement with each other and are treated independently.

\begin{figure} [hbt]
\includegraphics[width=14.0cm]{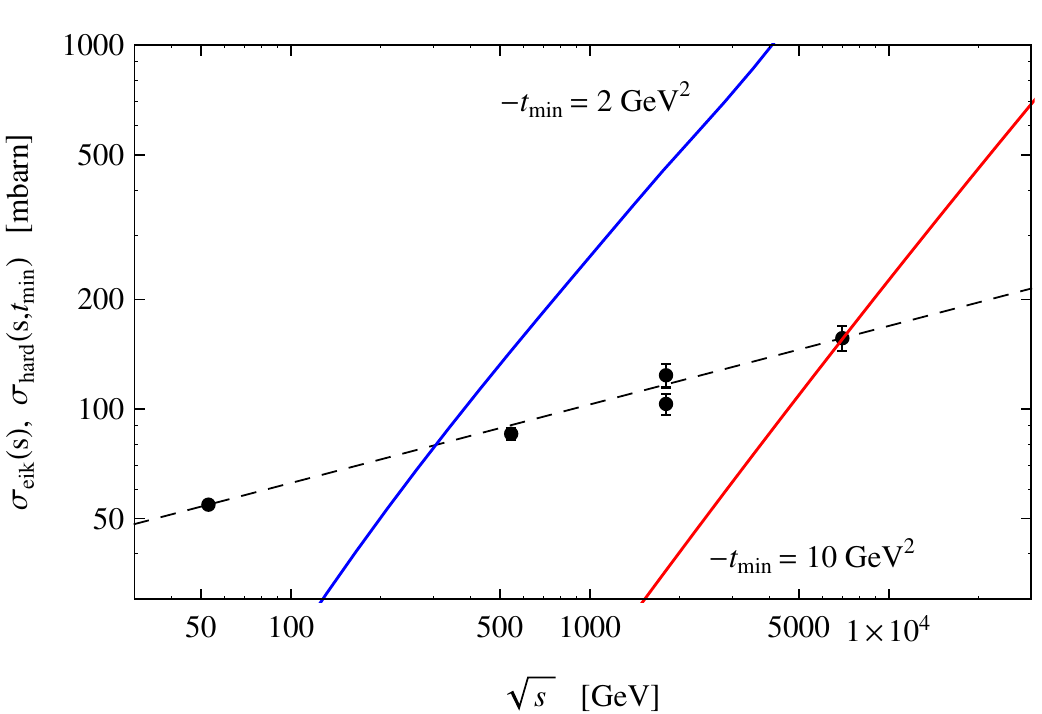}
\caption {\footnotesize 
The points are estimates of the quantity $\sigma_{\rm eik} (s)$
shown in fig.~\protect\ref{fig:error}, plotted as a function
of $\sqrt{s}$.
The thick lines are calculation 
of the cross section  for hard parton--parton scattering in $pp$ collisions 
in the kinematical regions
 $-t \ge 2$~GeV$^2$ and
 $-t \ge 10$~GeV$^2$. The calculation using LO cross parton
cross sections and the PDF's of MRSTW \protect\cite{Martin:2009iq}.
The dashed line, drawn to guide the eye
corresponds to $23~mb \times (s/{\rm GeV}^2)^{0.11}$.
\label{fig:sigma_hard} }
\end{figure}

\begin{figure} [hbt]
\includegraphics[width=14.0cm]{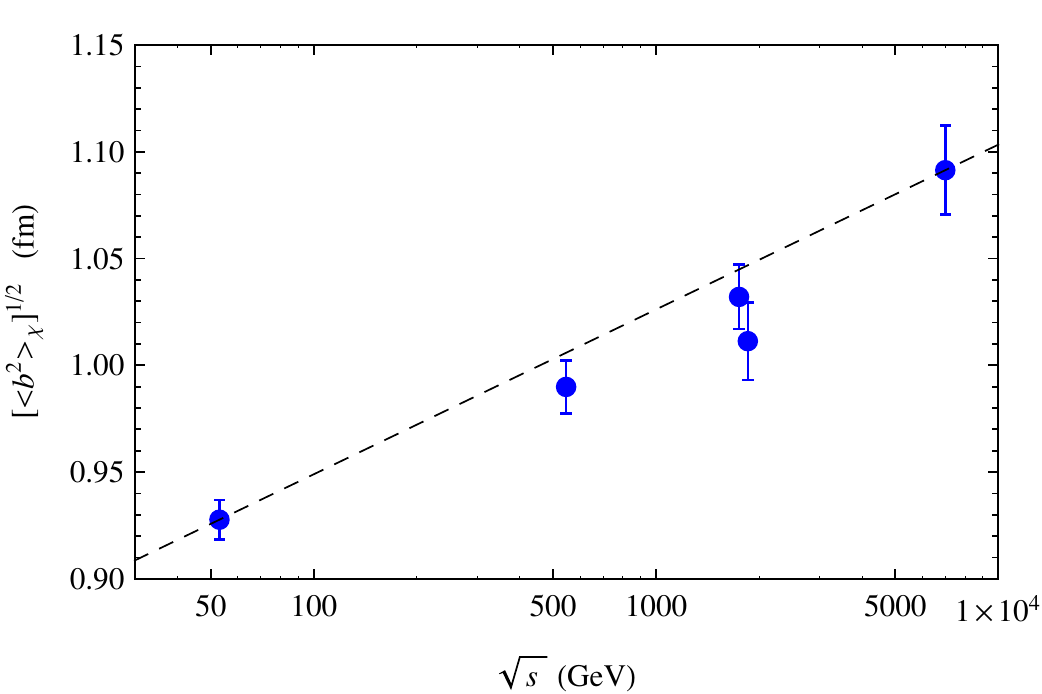}
\caption {\footnotesize 
Estimates of the quantity 
$\sqrt{\langle b^2 \rangle_\chi} = 4 \, r_0$ obtained from measurements 
of $\sigma_{\rm tot}$ and $B$ at the same c.m. energy.
The dashed line is a fit of form $a + b \; \ln s$ drawn to guide the eye.
\label{fig:b_average} }
\end{figure}

The energy dependences of the quantities $\sigma_{\rm eik}(s)$ and
 $\sqrt{\langle b^2 (s)\rangle_\chi}$ are shown in fig.~\ref{fig:sigma_hard} 
and~\ref{fig:b_average}.
The quantity $\sigma_{\rm eik}(s)$ grows approximately as a power law
($\sigma_{\rm eik}(s) \propto s^{0.11}$), 
while the width of the eikonal function grows
approximately logarithmically with $s$.

Evidence for the energy dependence of the width of the eikonal function
can also be seen from a study of the relation between 
the total cross section $\sigma_{\rm tot}$ and the slope
 $B_{\rm el}$, or equivalently between $\sigma_{\rm tot}$ 
and the ratio $\sigma_{\rm el}/\sigma_{\rm tot}$. 
These relations are not independent because the three quantities 
$\sigma_{\rm tot}$, $\sigma_{\rm el}$ and $B_{\rm el}$, 
in the approximation of a real elastic scattering amplitude 
(or $\rho^2 \ll 1$), are related by the optical theorem (\ref{eq:optical1}).

\begin{figure} [hbt]
\includegraphics[width=14.0cm]{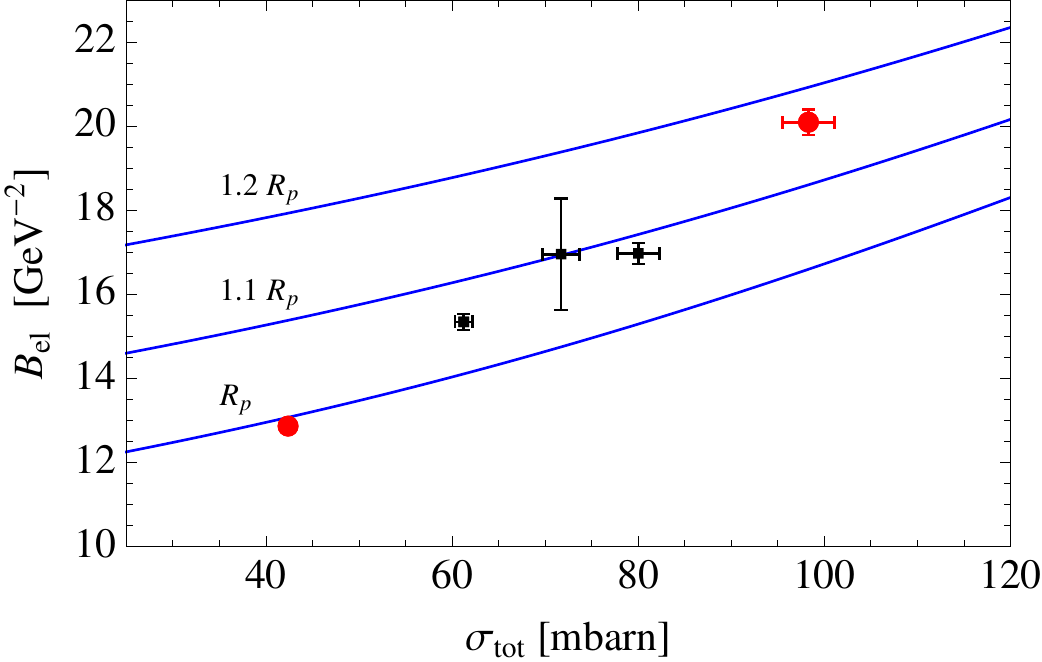}
\caption {\footnotesize 
Plot of the relation between $\sigma_{\rm tot}$ 
and the slope $B$. The (red) disks are measurements for
$pp$ scattering at ISR ($\sqrt{s} = 53$~GeV) and LHC
($\sqrt{s} = 7$~TeV); the (black) squares are measurements
for $\overline{p}p$ scattering at CERN ($\sqrt{s} = 546$~GeV)
and Fermilab ($\sqrt{s} = 1.8$~TeV).
The lines describe the relation when the eikonal function 
has the $b$ dependence of equation
(\ref{eq:App}) with the parameter
$r_0$ given by $1$, 1.1 and 1.2 times $R_p$ (with $R_p = 0.234$~fm obtained from
the electromagnetic form factor).
\label{fig:slope} }
\end{figure}

\begin{figure} [hbt]
\includegraphics[width=14.0cm]{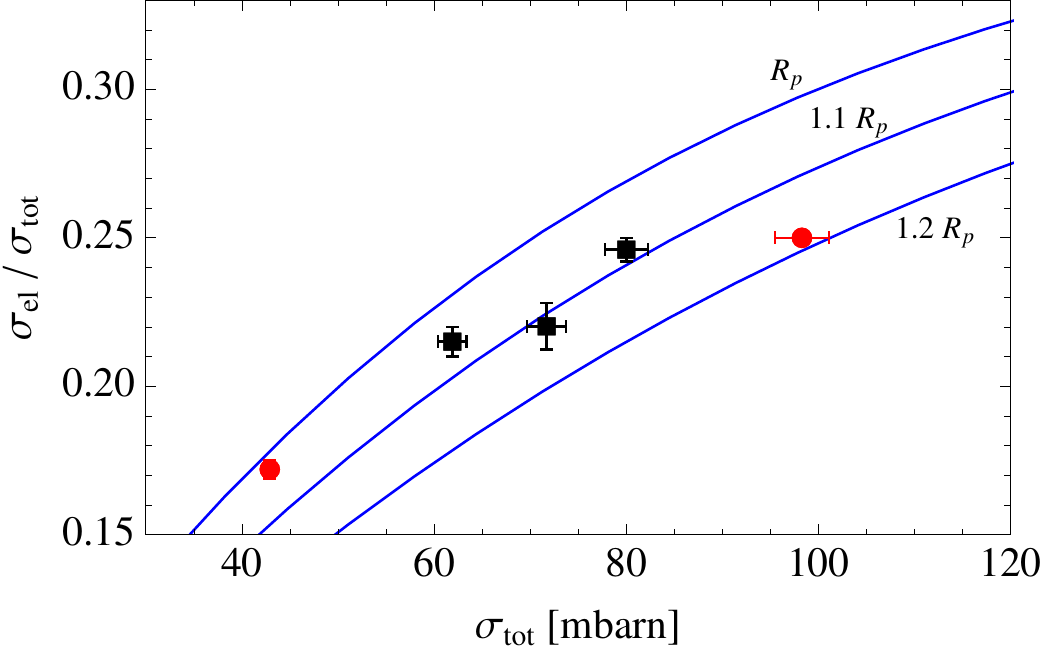}
\caption {\footnotesize
Plot of the relation $\sigma_{\rm el}/\sigma_{\rm tot}$
versus $\sigma_{\rm tot}$. The (red) disks are measurements for
$pp$ scattering at ISR ($\sqrt{s} = 53$~GeV) and LHC
($\sqrt{s} = 7$~TeV); the (black) squares are measurements
for $\overline{p}p$ scattering at CERN ($\sqrt{s} = 546$~GeV)
and Fermilab ($\sqrt{s} = 1.8$~TeV).
The lines describe the relation when the eikonal function
has the $b$ dependence of equation
(\ref{eq:App}) with the parameter
$r_0$ given by $1$, 1.1 and 1.2 times $R_p$ (with $R_p$ obtained from
the electromagnetic form factor).
\label{fig:ratio} }
\end{figure}

If one assumes that all the energy dependence is contained 
in the eikonal cross section, the darkening and broadening of the 
the opacity function with increasing $\sigma_{\rm eik}$ are not independent.
Accordingly, for any value of the total cross section (that is proportional
to the area under $\Gamma (b,s)$) one can compute 
the slope $B_{\rm el}$ (that measures the width of $\Gamma(b,s)$)
or the ratio $\sigma_{\rm el}/\sigma_{\rm tot} \propto \sigma_{\rm tot}/B_{\rm el}$.
The data, as illustrated in fig.~\ref{fig:slope} and fig.~\ref{fig:ratio},
indicate that the increase in the slope is faster,
and the increase in the ratio $\sigma_{\rm el}/\sigma_{\rm tot}$ is
slower than a prediction based on a constant
width for the eikonal function,
indicating that a broadening for 
increasing $\sqrt{s}$ is necessary.

\section{Single channel eikonal model}
\label{sec:single}
In a broad range of models \cite{Gaisser:1984pg,Pancheri:1985ix,Durand:1988cr},
the eikonal function $\chi(b,s)$ is interpreted as
proportional to $n(b,s)$, the average number of ``elementary interactions'' 
in a collision at impact parameter $b$ and c.m. energy $\sqrt{s}$:
\begin{equation}
 n(b,s) = 2 \, \chi(b,s) \;.
\label{eq:naive}
\end{equation}
This interpretation arises from the fact
that (using the approximation of a real eikonal function) one can rewrite 
the expression for 
the inelastic cross section (\ref{eq:siggen_inel}) in the form:
\begin{equation}
\sigma_{\rm inel} (s) = \int d^2b ~ \left [1 - e^{-2 \chi(b,s)} \right ] 
\end{equation}
The structure of this equation suggests that the quantity
$1-e^{-2 \, \chi(b,s)}$ has the physical meaning of the absorption probability 
in collisions at impact parameter $b$. If the absorption 
probability is understood as the probability 
of having at least one elementary interaction in the collision,
and in addition one assumes that the
fluctuations in the number of elementary interactions in the collisions
at fixed $b$ and $s$ have a Poissonian distribution,
one can interpret the quantity $2 \, \chi(b,s)$ as the 
average number of interactions, arriving to equation (\ref{eq:naive}).

Using the factorization of equation (\ref{eq:chi_decomp}),
one can see that from the assumption of equation (\ref{eq:naive})
one can conclude that both 
$\sigma_{\rm eik}(s)$ and $A_\chi(b,s)$ have simple and 
direct physical interpretations.
The quantity $\sigma_{\rm eik} (s)$ corresponds to the
parton--parton cross section at c.m. energy $\sqrt{s}$, and
$A_\chi(b,s)$ is interpreted as the overlap
of the spatial distributions of the interacting partons.

A fundamental problem for all models 
that attempt to describe the total cross sections
in terms of parton constituents in the colliding hadrons
is to compute in a consistent way a cross section that accounts for all 
elementary parton scatterings. 
In QCD a parton--parton cross section is
well defined and calculable only
for a momentum transfer sufficiently large.
For $Q^2 \to 0$ the perturbation theory expressions
for the cross sections diverge, but 
perturbation theory is not applicable,
and in fact also the identification of the partons in the 
colliding hadrons is uncertain. 

Several ideas have been put forward to compute
a finite cross section that accounts for all soft parton
interactions, but the problem remains open.
In this work we do not address this question but only try to
estimate a {\em lower limit} on this quantity.
The parton cross section can be decomposed into hard and soft components:
\begin{equation}
\sigma_{\rm parton} (s) = \sigma_{\rm hard} (s, |t_{\rm min}|) 
+ \sigma_{\rm soft} (s, |t_{\rm min}|)~
\end{equation}
where $\sigma_{\rm hard} (s,|t_{\rm min}|)$ accounts for
parton scatterings with momentum transfer $|t| \ge |t_{\rm min}|$, 
while $\sigma_{\rm soft}$ accounts for all other interactions.
It is obvious that $\sigma_{\rm hard} (s, |t_{\rm min}|)$ 
represents a {\em lower limit} to the total
parton cross section.
It is possible to choose the cutoff $|t_{\rm min}|$ sufficiently large,
so that the calculation of the hard cross section can be performed 
in perturbation theory 
convoluting the QCD parton scattering cross sections 
with the Parton Distribution Functions (PDF's) of the colliding hadrons.
An outline of such calculation
is described in appendix~A.
The results of a calculation 
performed
using the Leading Order PDF's of Martin, Roberts, Stirling, Thorne 
and Watt (MRSTW) \cite{Martin:2009iq} 
are shown in fig.~\ref{fig:sigma_hard} in the form 
of $\sigma_{\rm hard}(s, |t_{\rm min}|)$ plotted as a function of $\sqrt{s}$
for $|t_{\rm min}| = 2$ and 10~GeV$^2$.
In the figure the hard parton scattering cross section
is compared with $\sigma_{\rm eik} (s)$ as extracted from the data.
A comparison of the results shows that an identification
of $\sigma_{\rm parton}$ and $\sigma_{\rm eik} (s)$ are required 
in the framework discussed in this section is problematic.
The energy dependence of $\sigma_{\rm hard}(s, |t_{\rm min}|)$ 
for a fixed cutoff in momentum transfer
is much more rapid that the energy dependence of $\sigma_{\rm eik}(s)$.
At $\sqrt{s} = 53$, 546 and 7000~GeV
we have estimated $\sigma_{\rm eik} (s) \simeq 56$, 88 and 156~mb.
At the same energies for 
 $|t_{\rm min}| = 2$~GeV$^2$ one obtains $\sigma_{\rm hard} (s,t_{\rm min}) = 10$, 
120, and 1100~mb;
for $|t_{\rm min}| = 10$~GeV$^2$ 
$\sigma_{\rm hard} (s,t_{\rm min})$ is negligible at 
$\sqrt{s} = 53$~GeV and grows to 10~mb at 546 GeV and 120~mb at
7~TeV.

A possible solution to reconcile these results 
is to assume that the cutoff $|t_{\rm min}|$ is a function of $\sqrt{s}$.
With the LHC results this hypothesis becomes problematic.
At $\sqrt{s} = 7$~TeV the hard cross section calculated for
a cutoff as high as $|t_{\rm min}| = 10$~GeV$^2$ accounts for 
approximately 80\% of $\sigma_{\rm eik}(s)$, 
leaving little room for softer interactions.

An additional difficulty for the identification of
$2 \chi(b,s)$ with the average number of parton intercations $n(b,s)$
is the fact that the resulting number of parton
scatterings is smaller than what seems required by the observations
of the properties of the final state.

There is a broad consensus that a successful description 
of particle multiplicities and jet activities 
in high energy hadron collisions can only be achieved 
if the event generators incorporate a model for multiple parton scatterings.

A very attractive feature of models based on the ansatz of equation
(\ref{eq:naive}) is the possibility to introduce multiple--parton 
interactions in a simple and natural way.
Assuming that the fluctuations in the number of
elementary interactions are Poissonian
it follows that the probability of having exactly $k$ 
such interactions in a collisions (for fixed $b$ and $s$) is:
\begin{equation}
P_k (b,s) = 
e^{-n(b,s)} \; \frac{[n(b,s)]^k}{k!} 
\label{eq:Pk}
\end{equation}
(with $n(b,s) = 2 \, \chi(b,s)$).
The inelastic cross section is then decomposed
into the sum:
\begin{equation}
\sigma_{\rm inel}(s) = 
\sum_{k \ge 1} \; \int d^2b ~P_k (b,s) 
=
\sum_{k \ge 1} \sigma_k (s)\
\label{eq:sigma_partial}
\end{equation}
where each term corresponds to the cross section
for events with exactly $k$ elementary interactions.
The sum of all partial cross sections
weighted by the multiplicity $k$ yields:
\begin{equation}
\sum_k k~\sigma_k (s) = \sigma_{\rm eik}(s) \;,
\label{eq:sigma_parton}
\end{equation}
that is consistent with the interpretation of $\sigma_{\rm eik}(s)$
as the cross section for parton--parton scattering.
This last equation implies that the average number of 
parton interactions (for $\sqrt{s}$ fixed) is
\begin{equation}
\langle n(s) \rangle = \frac{\sigma_{\rm eik}(s)}{\sigma_{\rm inel}(s)} \;.
\label{eq:n_average}
\end{equation}

The first detailed Montecarlo model for multiple parton
interactions was constructed by Sj\"ostrand and van Zijl 
\cite{sjostrand-vanZijl}, on the basis of
these ideas. 
Essentially the same physical picture is present in other generators
that are now currently in use for the study of the LHC events
such as PYTHIA 8, Herwig++ or SHERPA.
These Montecarlo generators use the equivalent of 
equations (\ref{eq:Pk}) and (\ref{eq:sigma_partial}) to
estimate the probability of having a certain number of 
parton interactions in non diffractive interactions.
All these generators assume
that the parameters that describe the function $n(b,s)$ 
can be freely ``tuned'' to obtain a satisfactory agreement with the data.
The normalization of $n(b,s)$ determines the 
average multiplicity of inelastic events, while the 
shape in impact parameter of the function controls 
the width and form of the multiplicity distribution.

Using these algorithms the Montecarlo generators have been
able to reproduce with reasonable success the data. 
This purely phenomenological approach 
does not take into account the fact 
that in the framework used to derive the distribution
of parton interactions of equations (\ref{eq:Pk}) and
(\ref{eq:sigma_partial}) the quantity $n(b,s)$ is 
related to the eikonal function by equation (\ref{eq:naive}), 
and should therefore also be consistent with the measured 
total and elastic cross sections.

Reproducing the data taking into account this constraint 
is very difficult.
In the Montecarlo codes the increase in the average number of 
parton interactions per collision controls the growth of the 
average multiplicity in inelastic events.
When $\sqrt{s}$ grows from 53~GeV to 7~TeV
the density of charged particles in the central region 
(at pseudorapidity $\eta \simeq 0$) increases by a factor of approximately 3.1.
According to equation (\ref{eq:n_average}),
the average number of parton--parton interactions per inelastic collision
is also given by the ratio $\sigma_{\rm eik}(s)/\sigma_{\rm inel} (s)$.
When $\sqrt{s}$ grows from 53~GeV to 7~TeV
the ratio $\sigma_{\rm eik}(s)/\sigma_{\rm inel} (s)$,
increases only by a factor of order 1.4.
The relation between the number of parton interactions and the final
state multiplicity depends on many assumptions, but all existing 
generators require a faster growth of the number of elementary interactions
per collision to describe the data.

In the framework of a single channel eikonal model discussed in this section,
the shape in $b$ of the eikonal function $A_\chi(b,s)$
has a simple interpretation as the geometrical 
overlap of the spatial distributions of the interacting
partons in the colliding hadrons. 
According to this interpretation, 
and assuming that the spatial distribution of
quarks and gluons in a hadron is identical to
the electric charge distribution
(described by the electromagnetic form factor)
Durand and Pi suggested \cite{Durand:1988cr} the form (\ref{eq:App})
for the function $A_\chi(b)$ with its parameter 
to the value $r_0 = R_p$ (with $R_p = 0.234$~fm 
or $R_p^{-2} = 0.71$~GeV$^{2}$).

As discussed in the previous section,
the data on $pp$ scattering collected up to ISR energies are
in fact compatible with this hypothesis, but 
higher energy data require a broadening of the 
eikonal function with the growth of $\sqrt{s}$.
Durand and Pi, analysing the data collected at
the CERN $\overline{p}p$ collider \cite{Durand:1988ax}, 
have argued that up to ISR energies the main contribution to the 
parton--parton cross section is due to the scattering between 
valence quarks, while at higher energy gluon scattering becomes 
the dominant process. They also assumed that the spatial distributions
of valence quarks coincides with the charge distribution, while
gluons have a broader distribution. 
The same idea has been implemented by Block and Halzen 
in their ``Aspen model'' \cite{Block:2005pt}.

In these models the width of the eikonal function
becomes asymptotically constant at high energy
when the fraction of parton interactions due to gluon scattering
saturates. This conclusion is in tension with the
LHC results that indicate a continuous broadening of
$\chi(b,s)$ when $\sqrt{s}$ increases, 
as discussed in section~\ref{sec:opacity}.

A possibility is to assume that the distribution
in impact parameter space of a parton inside a proton
is $x$ dependent with a width that 
grows when $x$ becomes smaller.
Corke and Sj\"ostrand \cite{pythia1} 
have constructed a model that implements this idea.
In their model the gluon spatial distribution in a proton is gaussian
($\rho(r) \propto e^{-r^2/a^2(x)}$)
with a $x$ dependent width:
\begin{equation}
a(x) = a_0 \left ( 1 + a_1 \, \ln\frac{1}{x} \right )~. 
\end{equation}
The shape of $A_\chi(b,s)$ broadens 
with increasing c.m. energy, because the scattering of partons 
with smaller and smaller $x$ becomes possible.

\section{Multi--channel eikonal model}
\label{sec:multichannel}
The main motivation for the introduction
of a multi--channel eikonal is that in this framework
it is possible to include in a simple and 
consistent way the inelastic diffraction processes.
The fundamental idea was introduced by Good and Walker
\cite{Good-Walker} who observed that diffraction can be 
considered as analogous to the scattering of light 
from an absorbing screen that has aborption properties
that depend on the light polarization state.
The Good and Walker ansatz was later developed in by Miettinen
and Pumplin \cite{Miettinen:1978jb}, for a more recent 
reanalysis see also \cite{lipari-lusignoli}.

In the multi--channel eikonal framework the initial 
$pp$ state is considered as the linear combination of a 
complete set of orthogonal states $|\psi_j\rangle$:
\begin{equation}
\left | pp \right \rangle = \sum_j c_j \; \left | \psi_j \right \rangle 
\end{equation}
with $|c_j|^2 = p_j$ (from the normalization of the 
quantum states it follows that the sum of the 
coefficients $\{p_j\}$ is unity).
The states $|\psi_j\rangle$ are eigenstates of the scattering matrix,
so that when the state $j$ enters the collision 
(at impact parameter $b$ and c.m. energy $\sqrt{s}$) with unity 
amplitude, it emerges with amplitude $(1-G_j)$ and therefore
has absorption probability $1 - (1-G_j)^2$.

The opacity function can then be written as the sum:
\begin{equation}
\Gamma(b,s) = 
\sum_j p_j ~G_j (b,s) 
\label{eq:gamma-multi}
\end{equation}
In the following we will use the approximation that all
amplitudes $G_j$ are real and can take values only
in the intervals
\begin{equation}
0 \le G_j (b,s) \le 1~.
\label{eq:cond-G}
\end{equation}

In a multi--channel eikonal framework the inelastic cross section
is naturally decomposed into an absorption and a
diffraction component ($\sigma_{\rm inel} = \sigma_{\rm abs} + \sigma_{\rm diff}$).
Describing a cross section with the notation:
\begin{equation}
\sigma(s) = \int d^2b ~\tilde{\sigma}(b,s) \;,
\end{equation}
the quantity $\tilde{\sigma}_{\rm abs}$ can be expressed as:
\begin{equation}
\tilde{\sigma}_{\rm abs} (b,s) =
\sum_j p_j \; \left ( 1 - (1 - G_j)^2 \right )
= 2 \,\langle G(b,s) \rangle - \left \langle G^2(b,s) \right \rangle 
\label{eq:abs}
\end{equation}
In the last equation we have introduced the quantities
\begin{equation}
\langle G(b,s) \rangle = \sum_j p_j ~G_j(b,s) = \Gamma(b,s) \,
\label{eq:gaverage}
\end{equation}
and
\begin{equation}
\langle G^2(b,s) \rangle = \sum_j p_j ~G_j^2(b,s) \;.
\end{equation}
Subtracting the absorption from the inelastic cross section one
obtains for the diffractive cross section the expression:
\begin{equation}
\tilde{\sigma}_{\rm diff} (b,s) = 
\left \langle G^2(b,s) \right \rangle -\langle G (b,s) \rangle^2 ~.
\label{eq:sigdif}
\end{equation}
Equation (\ref{eq:sigdif}) shows that the diffractive cross section vanishes
when the dispersion of the $G_j(b,s)$ distribution vanishes.
This implies that all $G_j$ are identical,
and the multichannel framework reduces
to the single channel case discussed in the previous section.
The condition (\ref{eq:cond-G}) has the consequence that 
(for $b$ and $s$ fixed) the dispersion 
$\langle G^2 \rangle - \langle G\rangle^2$ 
has a maximum value that is obtained when the $G_j$ take values only
at the extremes of the allowed interval
(that is $G_j=1$ or $G_j=0$).
In this situation the opacity function $\Gamma(b,s)$ takes the form:
\begin{equation}
\Gamma = p_0 ~\delta[G] + p_1 ~\delta[G-1] \;.
\end{equation}
From (\ref{eq:gaverage}) one obtains 
$p_1 = \Gamma$ and $p_0 = (1-\Gamma)$, and therefore 
$\langle G^2 \rangle = \Gamma$
and $\tilde{\sigma}_{\rm diff} = \Gamma -\Gamma^2$.
In conclusion one has the bounds:
\begin{equation}
0 \le \tilde{\sigma}_{\rm diff} (b,s) \le \Gamma(b,s) -\Gamma^2(b,s) ~.
\label{eq:sdifa}
\end{equation}
From equations 
(\ref{eq:siggen_el}) and 
(\ref{eq:siggen_tot}) one also has
$\tilde{\sigma}_{\rm el}(b,s) = \Gamma^2(b,s)$ and
$\tilde{\sigma}_{\rm tot} (b,s)= 2 \,\Gamma(b,s)$, so 
that one can rewrite equation (\ref{eq:sdifa}) in the form:
\begin{equation}
\tilde{\sigma}_{\rm el}(b,s) + \tilde{\sigma}_{\rm diff} (b,s)\le 
\frac{\tilde{\sigma}_{\rm tot}(b,s)}{2} ~.
\label{eq:pumplin-bound}
\end{equation}
that after integration over impact parameter
gives the Miettinen--Pumplin's bound (\ref{eq:pumplin-bound1}).

Each function $G_j (b,s)$ can be written in the form: 
\begin{equation}
G_j (b,s) = 1 - e^{-\Omega_j(b,s)}
\end{equation}
introducing the partial eikonal functions
$\Omega_j(b,s)$ ($\Omega_j(b,s) \ge 0$).
The expression (\ref{eq:abs}) for the absorption cross section
can then be rewritten as:
\begin{equation}
\tilde{\sigma}_{\rm abs}(b,s) = 
\sum_j p_j ~\left [1 - e^{- 2 \,\Omega_j(b,s)} \right ]
\end{equation}
that, using arguments similar to those outlined in the previous section,
suggests to interpret the quantity 
$2 \, \Omega_j (b,s)$ as the average number of parton interactions
in a $pp$ collision (at impact parameter $b$ and c.m. energy $\sqrt{s}$)
when the initial state is the eigenstate $|\psi_j\rangle$:
\begin{equation}
n_j(b,s) = 2 \, \Omega_j(b,s) \;.
\label{eq:avomega}
\end{equation}
The average number of parton interactions per collision
can be obtained performing the summation:
\begin{equation}
n(b,s) 
= \sum_j p_j \; n_j (b,s) 
= 2 \; \sum_j p_j \; \Omega_j (b,s) 
= 2 \; \left \langle \Omega(b,s) \right \rangle \;.
\label{eq:multichannel}
\end{equation}

Equation (\ref{eq:multichannel}) 
is the multi--channel generalization of equation
(\ref{eq:naive}) that is recovered when the number of
distinct eigenstates is reduced to unity.
It is important to note that in this generalization 
equation (\ref{eq:naive}) becomes the limiting case of 
an inequality  that is valid in general:
\begin{equation}
n(b,s) \ge 2 \, \chi(b,s)~.
\label{eq:inequa}
\end{equation}
This result can be obtained 
rewriting equation (\ref{eq:gamma-multi}) in the form:
\begin{equation}
1 - e^{-\chi(b,s)} = 1- \sum_j p_j ~e^{-\Omega_j (b,s)}
\label{eq:mod-a}
\end{equation}
and using (\ref{eq:avomega}).

In the approximation that all partial eikonals $\Omega_j(b,s)$
have the same $b$ dependence one can write:
\begin{equation}
\Omega_j(b,s) = 
\left \langle \Omega(b,s) \right \rangle ~\alpha_j
=\frac{n(b,s)}{2} ~\alpha_j
\label{eq:alpha1}
\end{equation}
with $\alpha_j \ge 0$.
Passing to a continuous distribution 
one arrives to the expression \cite{Miettinen:1979ns,lipari-lusignoli}:
\begin{equation}
\Gamma(b,s) = \int_0^\infty d\alpha ~p(\alpha)
~ \left ( 1 - \exp \left [ 
-\frac{n(b,s)}{2} ~\alpha\right ] \right ) 
\label{eq:decompA}
\end{equation}
where the function $p(\alpha)$ satisfies the normalization condition:
\begin{equation}
\int_0^\infty d\alpha~p(\alpha) = 1
\end{equation}
and, because of equation (\ref{eq:alpha1}), also:
\begin{equation}
\int_0^\infty d\alpha~\alpha~p(\alpha) = 1~.
\label{eq:pa1}
\end{equation}

A multichannel eikonal model is now fully described by the
eikonal function $\chi(b,s)$ and the function $p(\alpha)$.
Note that the function $p(\alpha)$  is   equivalent
to the set of probabilities  $p_j$  in equation (\ref{eq:gamma-multi}),
the relation between these two  descriptions 
is discussed in the appendix~\ref{app:palpha}.
 
For a fixed eikonal, different 
choices for $p(\alpha)$ result in
different decompositions of the inelastic cross section
into absorption and diffraction components, and to a different
estimate of the quantity $n(b,s)$.

Miettinen and Thomas in \cite{Miettinen:1979ns} 
suggested for the function $p(\alpha)$ the form:
\begin{equation}
p(\alpha) = 
\frac{1}{w \; \Gamma_{\rm E} \left (\frac{1}{w} \right ) } 
~ \left ( \frac{\alpha}{w} \right )^{\frac{1}{w}-1}
 \exp \left [-\frac{\alpha }{w} \right ] 
\label{eq:p_model}
\end{equation}
(where $\Gamma_{\rm E}$ is the Euler Gamma function) that depends on
the real parameter $w \ge 0$.
It is simple to check that this form is normalized and has 
first and second moments: $\langle \alpha\rangle = 1$ and
$\langle \alpha^2\rangle = 1+w$.
In \cite{lipari-lusignoli}, unaware of the work of
Miettinen and Thomas, we independently 
suggested the form (\ref{eq:p_model}).
A discussion of the properties of the functional form (\ref{eq:p_model})
is presented in appendix~\ref{app:palpha}.

An important property of the parametrization (\ref{eq:p_model}) is that
when $w$ spans the interval $0 \le w < \infty$, the 
diffractive cross section spans all possible values
allowed by the Miettinen--Pumplin bound
($0 \le \sigma_{\rm diff} \le \sigma_{\rm tot}/2 - \sigma_{\rm el}$).

For $p(\alpha)$ of the form (\ref{eq:p_model}) 
it is also possible to perform analytically 
the integration in equation (\ref{eq:decompA}) 
and obtain a simple closed form expression for the 
relation between the opacity function
$\Gamma (b,s)$ (or the eikonal function $\chi(b,s)$)
 and the average number of parton interaction $n(b,s)$ that depends only on the 
parameter $w$:
\begin{equation}
n(b,s) = \frac{2 \; (e^{w \, \chi(b,s)}-1)}{w} 
= \frac{2}{w} \left ( \left [1 -\Gamma(b,s)\right ]^{-w} -1 \right )
~.
\label{eq:nnw}
\end{equation}
Expanding for small $w$ one finds:
\begin{equation}
n(b,s) = 2\, \chi(b,s) + \chi^2(b,s)\; w + \ldots \;
\label{eq:w-expa}
\end{equation}
The expansion is manifestly consistent with the inequality (\ref{eq:inequa}),
and in the limit of vanishing dispersion ($w \to 0$)
one returns to the single channel eikonal model result 
of equation (\ref{eq:naive}).
At the opposite limit, for
$w$ large, $n(b,s)$ grows exponentially with $w$.
The divergence of $n(b,s)$ for $w \to \infty$
can be readily understood qualitatively noting
that in this limit the $S$--matrix eigenstates 
are either completely trasparent or completely absorbed,
and $G_j=1$ implies $n_j \to \infty$.

The form (\ref{eq:p_model}) for $p(\alpha)$ also 
allows to obtain explicit expressions for 
 $\tilde{\sigma}_{\rm diff}(b,s)$
and $\tilde{\sigma}_{\rm abs}(b,s)$.
The quantity $\langle G^2\rangle$ can be calculated as:
\begin{eqnarray}
\langle G^2 \rangle & = & 
\int_0^\infty d\alpha ~p(\alpha) ~
\left (1 - \exp \left [ - \frac{n}{2} \, \alpha \right ] \right )^2
\nonumber \\ 
& ~ & \nonumber \\
& = &
1 + \left (1 + n(b,s) \, w \right )^{-\frac{1}{w}} 
- 2 \; \left (1 + \frac{n(b,s) \, w}{2} \right )^{-\frac{2}{w}} 
\label{eq:wdiff0}
\end{eqnarray}
and from equations (\ref{eq:sigdif}) and (\ref{eq:nnw}) one obtains:
\begin{equation}
\tilde{\sigma}_{\rm diff} (b,s) = 
[2 (1 -\Gamma )^{-w}-1]^{-1/w} - 1 + 2 \, \Gamma(b,s) - \Gamma^2 (b,s)~.
\label{eq:wdiff1}
\end{equation}
This expression can now be integrated
over $b$ to obtain the diffractive cross section as a function 
of the parameter $w$ using the opacity function obtained by
the elastic scattering measurement.

\begin{figure} [hbt]
\includegraphics[width=14.0cm]{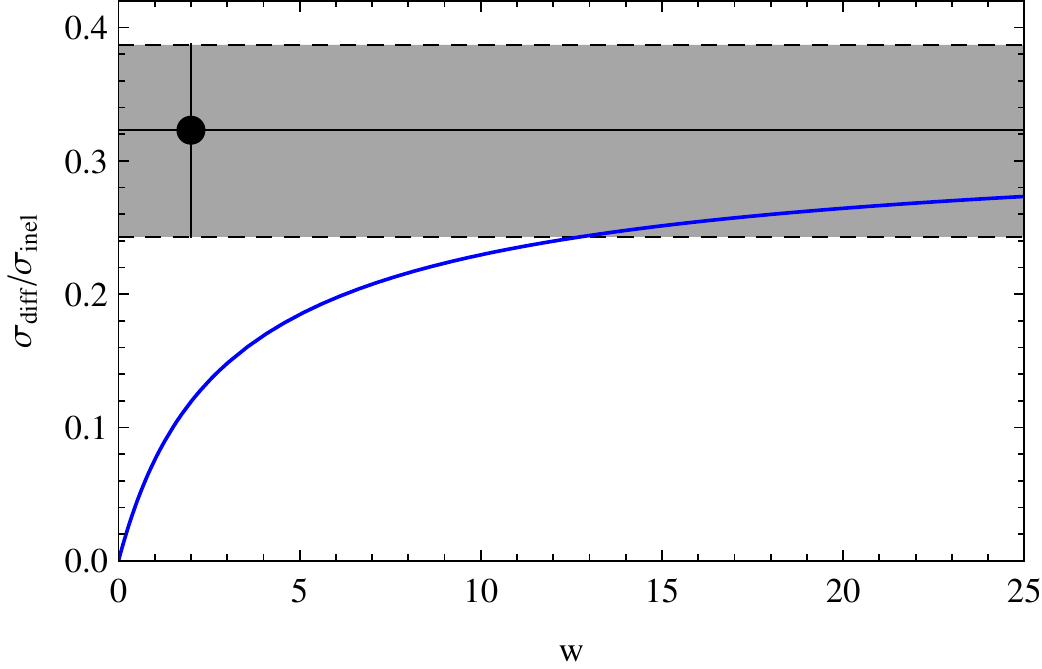}
\caption {\footnotesize The point with the error bar is the 
measurement of the ratio $\sigma_{\rm diff}/\sigma_{\rm tot}$ obtained
at $\sqrt{s} = 7$~TeV by the ALICE collaboration \protect\cite{alice-sigma}.
The line is the same ratio calculated in a multi--channel
eikonal framework based on equation (\ref{eq:p_model}) as a function
of the parameter $w$.
\label{fig:sdif} }
\end{figure}

The results of a numerical calculation of $\sigma_{\rm diff}$ at 
$\sqrt{s} = 7$~TeV using equation (\ref{eq:wdiff1}) 
are shown in fig.~\ref{fig:sdif}
in the form of the ratio $\sigma_{\rm diff} /\sigma_{\rm inel}$ plotted
as a function of $w$, and compared to the 
ratio measured by ALICE \cite{alice-sigma}:
$\sigma_{\rm diff}/\sigma_{\rm inel} = 0.32^{+0.064}_{-0.080}$. 
The diffractive cross sections vanishes for 
$w\to 0$ and grows monotonically with increasing $w$
going to the asymptotic value $\sigma_{\rm tot}/2 - \sigma_{\rm el}$
for $w \to \infty$. 

The experimental central value of the ratio 
$(\sigma_{\rm diff} + \sigma_{\rm el})/\sigma_{\rm tot}$ saturates the
Pumplin bound. This corresponds 
to the limiting case $w \to \infty$, and therefore
one can immediately see that from the study of the size of
the diffractive cross section one can only derive a lower limit for $w$.
At the 1~$\sigma$ level the limit is $w \ge 12.7$, 
and at 90\% C.L. is $w > 5.5$. 

In \cite{lipari-lusignoli} we argued that the data,
including the results obtained at the $\overline{p}p$ colliders
could be described, in the framework discussed in this section,
with a constant value $w \simeq 3$. The LHC data does not
support this hypothesis.

\begin{figure} [hbt]
\includegraphics[width=14.0cm]{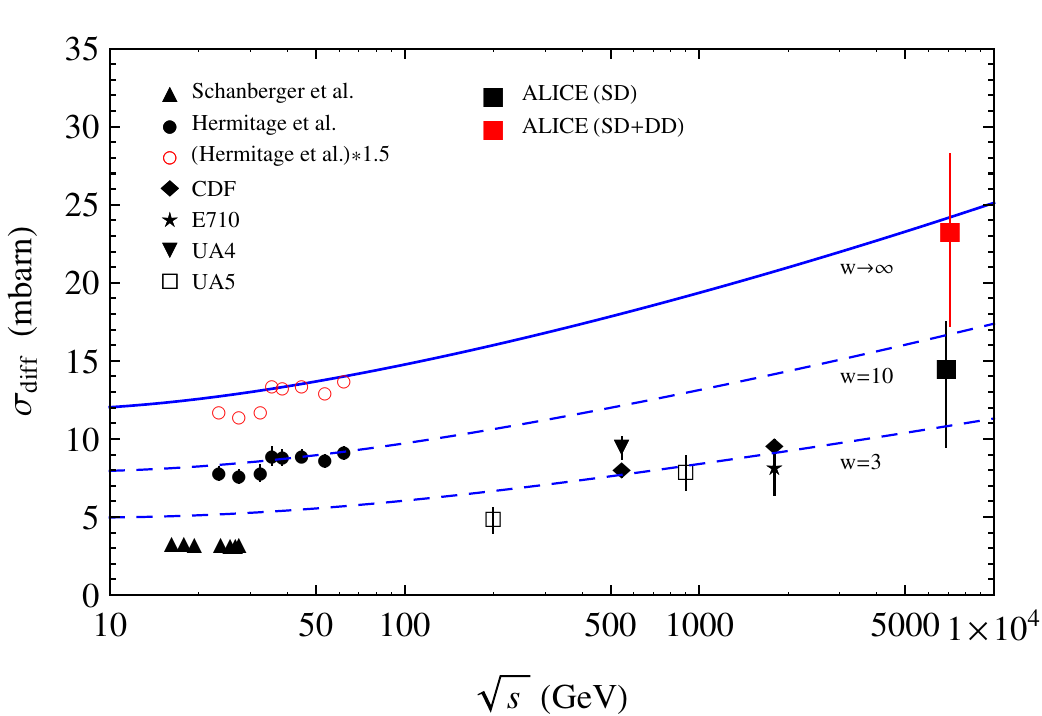}
\caption {\footnotesize 
The smaller points are measurements of the single diffractive cross sections
at ISR (Schamberger et al. \protect\cite{Schamberger:1975ea} and
Armitage et al. \protect\cite{Armitage:1981zp}), the $\overline{p}p$
colliders (UA4 \protect\cite{Bernard:1986yh},
UA5 \protect\cite{Ansorge:1986xq},
CDF \protect\cite{Abe:1993wu} and
E710 \protect\cite{Amos:1992jw}). The larger points
are measurement of the single and diffractive cross sections
at 7~TeV obtained combining the results of 
ALICE \protect\cite{alice-sigma} and TOTEM \protect\cite{Antchev:2011vs}.
The lines calculation of the diffractive cross
sections described in the main text.
\label{fig:pldif} }
\end{figure}

The situation for the diffractive cross section as a function
of energy is summarized in fig.~\ref{fig:pldif} that shows
the measurements of the diffractive 
cross section obtained at 7~TeV at LHC together with 
data on single diffraction
obtained at ISR (where the 
results of Schamberger et al.\cite{Schamberger:1975ea} and
Armitage et al. \cite{Armitage:1981zp} are in poor agreement with each other)
and at $\overline{p}p$ collider energies
(UA4 \cite{Bernard:1986yh},
UA5 \cite{Ansorge:1986xq},
CDF \cite{Abe:1993wu} and
E710 \cite{Amos:1992jw}).
The ISR results refer to the kinematical region
$M^2/s < 0.1$, while the collider data refer to $M^2/s < 0.05$.
To compute the diffractive cross sections shown in the figure 
we have used the model described above,
based on equation (\ref{eq:p_model}) with the
eikonal function parametrized 
as in equation (\ref{eq:chi_decomp}) with $A_\chi(b,s)$ of the form
(\ref{eq:App}). 
The parameters $r_0(s)$ and $\sigma_{\rm eik}(s)$ 
have been calculated from the values of $\sigma_{\rm tot}(s)$ 
and $\sigma_{\rm el}(s)$
estimated with the fits to the total cross section
from the PDG--2010 \cite{pdg-2010} and of the elastic cross section
from \cite{Antchev:2011vs}. 

The prediction for the diffractive cross section we are discussing
here, based on the Good and Walker ansatz, refers
to the sum of the single and double diffraction cross section,
on the other hand most of the data is only for single diffraction processes,
because of the experimental difficulty in separating double diffraction
from non--diffractive interactions. This  introduces  ambiguity
in the interpretation.
Inspecting fig.~\ref{fig:pldif}  one can see that a  consistent 
interpretation of the data is  not easy.
An interesting  speculation is that, allowing
for systematic errors, and including a double diffraction
cross section of order $\sigma_{\rm DD}\simeq \sigma_{\rm SD}/2$,
the Miettinen--Pumplin bound is  saturated at all energies
for $s \gg 4 \, m_p^2$.

In the simple model discussed here 
the single parameter $w$ completely determines the structure
of the multi--channel eikonal, and allows to obtain the
quantity $n(b,s)$ from the eikonal function, and so to make contact
with a partonic description of the interaction.
The relation between $\chi(b,s)$ and $n(b,s)$ was given in 
equation (\ref{eq:nnw}).

As discussed above the quantity $n(b,s)$ grows monotonically
with $w$, diverging exponentially for $w$ large. 
This is problematic in a situation where the data suggests that
$w$ is in fact large. A divergence of 
of $n(b,s)$ is not an immediately fatal problem since it does not
imply the divergence of directly observable quantities,
and in fact the QCD parton cross section are divergent in perturbation theory.

Equation (\ref{eq:nnw}) allows to compute the shape of the function
$n(b,s)$ that has the physical meaning of the overlap of 
the hadronic matter distributions in the colliding hadrons.
It is simple to see that in a multichannel eikonal framework
the width of $n(b,s)$ is always narrower
than the width of $\chi(b,s)$. In the scheme discussed here
the width decreases monotonically with the increase of $w$.

Miettinen and Thomas \cite{Miettinen:1979ns}, 
have estimated the spatial distribution of hadronic matter in the proton
from the width of the eikonal function 
measured from the data on $pp$ scattering at ISR.
In a single channel eikonal framework the width of the eikonal
$\langle b^2 \rangle_\chi \simeq 0.93$~fm suggests that hadronic matter
has the same distribution of electric charge 
(see discussion in sec.~\ref{sec:opacity}), but
in multi--channel eikonal one has to conclude that 
hadronic matter has a significantly more compact distribution.

In the higher energy LHC data the eikonal function becomes
broader, but also the parameter $w$ seems to grow, indicating
that the interacting hadronic matter becomes more compact
with increasing energy. This statement is in fact opposite to what one
obtains in a single channel framework.

\section{Discussion}
\label{sec:discussion}
It is natural to expect that the same fundamental dynamics determines the
energy dependence of the total, elastic and diffractive
cross section in $pp$ interactions and also controls the main
properties of particle production in inelastic
interactions such as multiplicity distributions, particle composition and
inclusive energy spectra. 
From this idea it follows that it is not only desirable but 
necessary to construct a comprehensive and consistent theory 
that predicts the energy dependence 
of $\sigma_{\rm tot}$, $\sigma_{\rm el}$ and $\sigma_{\rm diff}$
and at the same time is the basis for a description 
of the final state in the collisions.
It is also natural to expect that this comprehensive theory will
be based on the partonic structure of colliding hadrons.
Single and multi--channel eikonal models 
seems the most promising 
approaches for the construction of a parton--based model.
In this work we have tried
to interpret the  cross sections  measurements recently obtained at LHC 
in these frameworks.

The eikonal function $\chi(b,s)$ completely determines
(and is completely determined by) the elastic scattering amplitude.
Using the approximation that the amplitude is purely imaginary
the eikonal function is real and can be calculated from 
the differential elastic cross section if 
the data cover a sufficiently broad range in momentum transfer $t$. 
In this work we have used the TOTEM data at 7~TeV 
to estimate the eikonal function.
Comparing to the data on $pp$ scattering at 
lower energy,  the eikonal function becomes larger
and broader. The normalization of the eikonal function
can be expressed as an eikonal cross section $\sigma_{\rm eik} (s)$
that grows from $55\pm 1$~mbarn at $\sqrt{s} = 53$~GeV,
to $156 \pm 14$~mbarn at $\sqrt{s} = 7$~TeV. 
The width of the eikonal 
(as measured by $\langle b^2 \rangle_\chi$) 
grows from $0.86 \pm 0.02$~fm to $1.19 \pm 0.04$~fm. 

In single channel eikonal models the eikonal cross section $\sigma_{\rm eik}(s)$
is interpreted as the total parton--parton cross section, while the 
shape in $b$ of $\chi(b,s)$ is the convolution of the spatial
distributions of the interacting partons.
A $\sqrt{s} = 7$~TeV a cross section of order 150~mbarn corresponds,
in a standard perturbative QCD calculation, to the cross section
for (semi)--hard hadronic jet--pair production for $p_\perp^2 \simeq 5$~GeV$^2$,
and one expects that the total parton--cross section should be 
significantly larger. The identification of the parton
and eikonal cross section 
that is the key element in a single channel eikonal framework is 
therefore problematic since $\sigma_{\rm parton} > \sigma_{\rm eik}$.

A possible solution is to introduce some modification of the 
standard calculation to reduce the parton cross section.
This idea however encounters other difficulties when
the eikonal framework is used as the basis 
for the multi--parton structure of inelastic events,
a method that is in fact adopted 
in most of the Montecarlo event generator used to model the interactions
at LHC.
The general idea is that 
the complexity and average multiplicity of the final state
grow with c.m. energy because 
the number of parton interactions per inelastic collision increases
with $\sqrt{s}$.
In a single channel eikonal 
framework the ratio $\sigma_{\rm eik}(s)/\sigma_{\rm inel}(s)$ 
has the physical meaning of the average number of parton
interactions per collision.
From the data  one finds that the 
ratio $\sigma_{\rm eik}(s)/\sigma_{\rm inel}(s)$ 
increases from a value $1.55 \pm 0.03$ at $\sqrt{s} = 53$~GeV
to $2.1 \pm 0.2$ at 7~TeV. 
For the same c.m. energies 
the density of charged particles 
in the central region ($\left .dN/d\eta \right |_{\eta = 0}$)
grows significantly faster,
from approximately 1.4 at $\sqrt{s} = 53$~GeV to 6.1 at
7~TeV. 
For the event generators it is very difficult
to reproduce this increase in  average multiplicity 
respecting the constraint  that the average number of
interactions  grows $\propto \sigma_{\rm eik}/\sigma_{\rm inel}$.

In view of these difficulties, we conclude 
that a single channel eikonal framework is not viable,
and that it is necessary to consider a multi--channel model.
This more complex framework  is in fact desirable 
also  because  it has the very important 
theoretical  merit to allow  the inclusion 
of inelastic diffraction in a consistent way.
In a multi--channel framework the eikonal cross section 
is not identified with the parton cross section
but represents only a lower limit: $\sigma_{\rm eik} \le \sigma_{\rm parton}$.
This opens the possibility to solve the problems 
outlined above.

In a multi--channel framework the initial state is decomposed
in a set of eigenstates that have distinct 
transmission amplitudes ($1-G_j$).
The model is fully defined when all these transmission amplitudes
are specified, respecting the condition that the weighted sum of 
the amplitudes reproduces the opacity function
(see equation (\ref{eq:gamma-multi})).
The diffractive cross section emerges when the distinct transmission
amplitudes are different from each other,
and is proportional 
to the dispersion of the distribution of the transition amplitudes.
From the assumption that each amplitude is in the interval $0 \le G_j \le 1$
it follows that the dispersion  has a  finite upper  limit,
and therefore the diffractive cross section
 has a maximum theoretically
allowed value, the so called Miettinen Pumplin bound:
$\sigma_{\rm diff} \le \sigma_{\rm tot}/2 - \sigma_{\rm el}$.
The inequality is saturated
when the transmission amplitudes take only the
extreme values (zero or unity), in other words when all
eigenstates in the Good--Walker decomposition 
are either completely absorbed or perfectly transparent. 

The parton cross section is also calculable from the structure
of the multi--channel model, and is related to the size of the
diffractive cross section. It takes its minimum value
($\sigma_{\rm parton} = \sigma_{\rm eik}$) when the diffractive cross
section vanishes, and grows monotonically with $\sigma_{\rm diff}$,
diverging when the diffractive cross section approaches its maximum
theoretically allowed value.
It is remarkable that the combination of the TOTEM and ALICE data 
at $\sqrt{s} = 7$~TeV is consistent with the saturation 
of the Miettinen Pumplin bound. This implies 
that the parton cross section is very large,
and in fact divergent if the bound is  saturated.

In this work we have discussed a ``minimum model'' 
to describe the multi--channel eikonal  
already  discussed in the literature \cite{Miettinen:1979ns,lipari-lusignoli}.
The  model contains a
single parameter  ($w \ge 0$) that is proportional 
to  the dispersion of the eigenvalues  of the partial eikonals $\Omega_j$.
Using this model one obtains a lower limit on $w$
that corresponds to a lower limit on the parton cross section
of the order of $6 \times 10^{6}$~mbarn (at 95\% C.L.).

This result suggests that the parton cross section is 
in fact divergent.
The divergence is not catastrophic, because it 
appears in the negative argument of an exponential, and it implies
that a set of scattering eigenstates is absorbed with 
unit probability. 

\vspace{0.3 cm}
\noindent{\bf Acknowlegments.}  We  are  grateful to  
Lia Pancheri, Yogi Srivastava and Daniel Fagundes
for  many discussions  on the problems discussed in this work.

\clearpage

\appendix

\section{Parton interaction cross sections}
\label{app:partons}
In the theoretical framework we are considering, one introduces
a ``parton cross section'' $\sigma_{\rm parton} (s)$
 that is related to the total number of elementary interactions
in $pp$ collisions. One must have that $\sigma_{\rm parton}(s)$
is larger than $\sigma_{\rm inel}(s)$ because in general in 
an inelastic $pp$ scattering one must have at least one such elementary
interaction, and in fact the ratio $\sigma_{\rm parton}(s)/\sigma_{\rm inel}(s)$
has the physical meaning of the average number of
elementary interactions per inelastic event.
The calculation of the quantity $\sigma_{\rm parton}(s)$ 
is a non trivial problem, because the parton--parton cross sections
(calculated at tree level) diverge for low momentum transfer, when
in fact perturbation theory fails.
Several authors have dealt with this problem decomposing 
$\sigma_{\rm parton}(s)$ in a ``hard'' (calculable in perturbation 
theory) and a ``soft'' component that is 
estimated with different methods.

In this appendix we want to evaluate the cross section for
parton--parton hard scattering in $pp$ collisions,
in the region where one can be confident that 
a perturbative calculation is valid.
For example, the differential cross section 
for elastic gluon--gluon scattering 
can be calculated in perturbation theory
convoluting the gluon distribution functions
with the elementary cross section for gluon--gluon scattering
 with the result:
\begin{equation}
\frac{d^3 \sigma_{gg \to gg} }{dt \, dx_1 \, dx_2}
(t, x_1 , x_2; \sqrt{s})
= 
~f_g (x_1, |t|) 
~
f_g (x_2, |t|) 
~\frac{d \hat{\sigma}_{gg\to gg}}{dt} (t, \hat{s})
\label{eq:gg_scattering}
\end{equation}
where $x_1$ and $x_2$ are the fractional momenta carried by the
gluons in the colliding protons and $\hat{s} = s \, x_1 \, x_2$.
It is then possible to 
obtain the cross section $\sigma_{gg}(s,|t_{\rm min}|)$
for all gluon--gluon scatterings 
with momentum transfer larger than $-t_{\rm min}$,
integrating the expression above over the appropriate kinematical range.
\begin{eqnarray}
\sigma_{gg} (s, t_{\rm min}) & = & 
\int dx_1 ~\int dx_2 ~\int d|t|
~\theta \left ( x_1 \, x_2 - \frac{2 |t|}{s} \right )
~\theta ( |t| - |t_{\rm min}|) 
\times ~~~~
\nonumber 
\\[0.2 cm] 
& ~ & 
~~~~~~~~~f_g(x_1, |t|) \; f_g (x_2, |t|)
~\frac{d\hat{\sigma}_{gg}}{dt} (t,\hat{s})
\label{eq:sigmagg}
\end{eqnarray}

Similarly one can compute cross sections for quark--gluon and quark--quark
scattering. Combining these results one can obtain the
hard cross section as the combination of $gg$, $qg$ and $qq$ scatterings:
\begin{equation}
\sigma_{\rm hard} (s,|t_{\rm min}|) = 
\sigma_{gg} +
\sigma_{qg} +
\sigma_{qq} 
\end{equation}

The cross sections for hard parton scattering 
scale approximately as 
$|t_{\rm min}|^{-1}$, reflecting
the divergence ($\propto t^{-2}$) of the 
parton--parton differential elastic cross section, and grows
rapidly with $s$, because with increasing energy 
the hard scattering of partons 
with lower $x$ 
($x_1 \, x_2 \ge 2 \, |t_{\rm min}|/s$) 
becomes kinematically possible.

To obtain numerical estimates of $\sigma_{\rm hard} (s,|t_{\rm min}|)$
we have used the Leading Order PDF's of 
Martin, Roberts, Stirling, 
Thorne and Watt (MRSTW) \cite{Martin:2009iq}.
The results of the calculation 
are shown in fig.~\ref{fig:sigma_hard} 
as a function of the $|t_{\rm min}|$ cut for fixed values of 
$\sqrt{s} = 53$, 546 and 7000~GeV.

At the LHC energy ($\sqrt{s} = 7$~TeV), the 
hard parton scattering cross section is $\sigma_{\rm hard} = 1465$~mbarn
for $|t_{\rm min}| = 2$~GeV$^2$, decreasing to 162~mbarn for
$|t_{\rm min}| = 10$~GeV$^2$.

For a qualitative understanding one can observe
that using some approximations it is possible to obtain 
an analytic expressions for the cross sections for 
hard parton scattering.

The elementary differential cross section
for gluon--gluon scattering 
has the form:
\begin{eqnarray}
\frac{d\hat{\sigma}_{gg}}{dt} (t, \hat{s}) & = &
\frac{9\,\pi}{2} \, \frac{\alpha_s^2(|t|)}{\hat{s}^2}
\; \left [3 
- \frac{ t \, u}{\hat{s}^2} 
- \frac{ \hat{s} \, u}{t^2} 
- \frac{ \hat{s} \, t}{u^2} \right ] 
\nonumber 
\\[0.2 cm]
& \simeq & 
\frac{9 \, \pi}{2} \, \frac{\alpha_s^2(|t|)}{t^2} ~\left [1 + 
O \left ( \frac{t}{\hat{s}}\right )
\right ]
 \end{eqnarray}
that diverges as $t^{-2}$ for $t\to 0$.
The integration over $t$ can be performed 
analytically if one neglects the scale dependence of $\alpha_s$.
The result is:
\begin{eqnarray}
\hat{\sigma}_{gg} (\hat{s},|t_{\rm min}|) & = &
\frac{9\pi}{2} \, \frac{\alpha_s^2}{|t_{\rm min}|} 
~F_{\rm kin} \left ( \frac{2 |t_{\rm min}|}{\hat{s}} \right ) 
\nonumber \\
& = & 
\hat{\sigma}_{gg}^{\rm asy} (|t_{\rm min}|) 
~F_{\rm kin} \left ( \frac{2 |t_{\rm min}|}{\hat{s}} \right ) ~.
\end{eqnarray}
The function $F_{\rm kin}(\hat{\tau})$ is a kinematical
suppression factor that takes into account the
available phase space (with $\hat{\tau} = 2 |t_{\rm min}|/\hat{s}$).
It vanishes at the threshold
($\hat{\tau} = 1$) and is equal to unity for $\hat{\tau} = 0$
(that corresponds to $\hat {s} \to \infty$).
The exact expression for $F_{\rm kin}(\hat{\tau})$ is:
\begin{equation}
F_{\rm kin}(\hat{\tau}) = 1 -
\hat{\tau} \left [\frac{(20 - 106 \, \hat{\tau} 
+ 42 \,\hat{\tau}^2 - 5 \,\hat{\tau}^3 + \hat{\tau}^4)}{48 \, (2-\hat{\tau})}
+ \frac{1}{2}\; \log \left ( \frac{2}{\hat{\tau}} -1 \right ) \right ]
\end{equation}
For numerical purposes this can be approximated with the simple
form: $f(\hat{\tau}) \simeq (1-\hat{\tau})$ that has 
correct values for $\hat{\tau} = 0$ (large $\hat{s}$ limit) and
 $\hat{\tau} = 1$ (threshold).

Neglecting the scale dependence of the PDF's, 
one can perform the $t$ integration in (\ref{eq:sigmagg})
and the $gg$ scattering cross section results factorized in the form:
\begin{equation}
\sigma_{gg} (s, t_{\rm min}) \simeq 
\hat{\sigma}_{gg}^{\rm asy} (|t_{\rm min}|) \times
C_{gg} (\tau) 
\end{equation}
(with $\tau = 2 |t_{\rm min}|/s$)
that is the product of the (asymptotic) elementary gluon--gluon
cross section with the convolution of the gluons PDF's $C_{gg} (\tau)$ 
that computes the the number of gluon pairs
above threshold (that is with $x_1 x_2 > \tau$),
with the function $F_{\rm kin} [\hat{\tau} = \tau/(x_1 \,x_2)]$
that takes into account the phase space available for the 
scattering: 
\begin{eqnarray}
C_{gg} (\tau) & = &
\int dx_1 \int dx_2 
~\theta \left ( x_1 \, x_2 - \frac{2 |t_{\rm min}|}{s} \right )
\times ~~~~
\nonumber 
\\[0.2 cm]
& ~ & 
~~~~f_g(x_1, |t_{\rm min}|) \; f_g (x_2, |t_{\rm min}|)
~F_{\rm kin} \left ( \frac{\tau}{x_1 x_2} \right )
\nonumber 
\end{eqnarray}
If the gluon PDF's are approximated with the simple form
$f_g(x) = K_g/x^{1+\varepsilon}$, 
the convolution factor can be calculated explicitely:
\begin{equation}
C_{gg} (\tau) = K_g^2 \left [ -\frac{\tau^{-\varepsilon} \; \log \tau}
{\varepsilon \, (1 + \varepsilon)} 
-\frac{1 + 2 \, \varepsilon + \varepsilon^2 \, (1 - \tau) 
- \tau^{-\varepsilon} \, (1 - 2 \, \varepsilon) }
{\varepsilon^2 (1 + \varepsilon)^2}
 \right ]
\end{equation}
In the limit of $\varepsilon \to 0$ the convolution factor becomes:
\begin{equation}
C_{gg} (\tau) \simeq K_g^2 \; 
\left [
\frac{\log^2 \tau}{2}
+\log \tau +1 - \tau \right ]\;.
\end{equation}
In the limit of high energy ($s/(2 |t_{\rm min}|) \gg 1$, or $\tau \to 0$)
their asymptotic behavior is:
\begin{equation}
C_{gg} \simeq 
K_g^2 ~\frac{1}{\varepsilon} \;
\left (\frac{s}{2 \, |t_{\rm min}|} \right )^\varepsilon
\log \left (\frac{s}{2 \, |t_{\rm min}|} \right )
\end{equation}
or for $\varepsilon \to 0$:
\begin{equation}
C_{gg} \simeq K_g^2 \; \frac{1}{2} ~
\log^2 \left ( \frac{s}{2 \, |t_{\rm min}|} \right )
\end{equation}

The cross sections for quark--gluon and quark--quark 
scattering can be written as similar decompositions,
noting that the asymptotic behavior of elementary cross sections
differ by simple color factors:
\begin{equation}
 \hat{\sigma}_{gg}^{\rm asy} 
= 
\frac{9}{4} \; \hat{\sigma}_{qg}^{\rm asy} =
\left (\frac{9}{4} \right )^2 \; \hat{\sigma}_{qq}^{\rm asy} ~.
\end{equation}

\section{The functions $p(\alpha)$ and $p_G (G)$}
\label{app:palpha}

In a multi--channel eikonal model
the opacity function $\Gamma(b,s)$ is decomposed into 
partial components as:
\begin{equation}
\Gamma = \sum_j p_j ~G_j
\label{eq:multi-a}
\end{equation}
(where we have left implicit the $b$ and $s$ dependence).
The partial opacities $G_j(b,s)$ are assumed to be real
and in the interval $[0,1]$, therefore in the most general case one 
has to consider a continuous infinity of states that can be labeled 
with the eigenvalue $G$. 
The decomposition (\ref{eq:multi-a})
can then be written as the integral:
\begin{equation}
\Gamma = \int_0^{1} dG~ p_G (G) ~G = \left \langle G \right \rangle
\label{eq:decompG}
\end{equation}
(where the probability distribution $p_G (G)$ is normalized).
The discrete case can be recovered using for the function $p_G(G)$:
\begin{equation}
p_G (G) = \sum_j p_j ~\delta [G - G_j]~.
\end{equation}

One can relate $G$ to the parameter $\alpha$ with the relation
\begin{equation}
G = 1 - e^{-\Omega} = 1 - e^{-\langle \Omega \rangle \, \alpha}~.
\end{equation}
Since $\langle \Omega \rangle$ is positive, 
and $G$ can vary in the interval $[0,1]$,
the quantity $\alpha$
takes values in the interval $ 0 \le \alpha < \infty$.
The decomposition of the profile function 
can then be rewritten in the form:
\begin{equation}
\Gamma = \int_0^\infty d\alpha ~p(\alpha)
~ \left ( 1 - e^{-\langle \Omega \rangle \; \alpha} \right ) ~.
\label{eq:decompAa}
\end{equation}
The probability distribution of $\alpha$ is normalized
to unity and satisfies the constraint:
\begin{equation}
\int_0^\infty d\alpha ~\alpha ~p(\alpha) = 1~.
\label{eq:alpha2}
\end{equation}
As discussed in the main text,
one can make the interpretation:
$\langle \Omega \rangle = n/2$ where $n$ is the average number of
parton interactions.

The two decompositions of the opacity functions 
in equations (\ref{eq:decompG}) and (\ref{eq:decompA}) 
are equivalent. There is 
a one--to--one correspondence between $G$ and $\alpha$,
and one can obtain $p(\alpha)$ fron $p_G(G)$ or viceversa.
Equation (\ref{eq:decompAa}) determines implicitely
$\langle \Omega \rangle$ from $\Gamma$ 
(or viceversa $\Gamma$ from $\langle \Omega \rangle$)
if the function $p(\alpha)$ (or $p_G(G)$) is known.

The important quantity $\langle G^2 \rangle$
that enters the expression for the diffractive and absorption cross
sections (see equations (\ref{eq:abs}) and (\ref{eq:sigdif}))
can be calculated in the two decompositions as:
\begin{eqnarray}
\langle G^2 \rangle
 & = & \int_0^1 dG ~p_G (G) ~G^2 \\
~& ~& \nonumber \\
\langle G^2 \rangle
& =& 
\int_0^\infty d\alpha ~p(\alpha)~\left (1 -e^{-\langle \Omega \rangle \; \alpha} \right )^2 
\end{eqnarray}

In this work we have used for $p(\alpha)$ the form 
\begin{equation}
p(\alpha) = 
\frac{1}{w \; \Gamma_{\rm E} \left (\frac{1}{w} \right ) } 
~ \left ( \frac{\alpha}{w} \right )^{\frac{1}{w}-1}
 \exp \left [-\frac{\alpha }{w} \right ] 
\label{eq:pa_model}
\end{equation}
(aso given in the main text in equation (\ref{eq:p_model}))
that depends on the parameter $w \ge 0$.
For $n$ integer one has:
\begin{equation}
\left \langle \alpha^n \right \rangle = \int_0^\infty d\alpha
\; p(\alpha) \; \alpha^n = (1+w) \, (1+ 2 w) \ldots [1+ (n-1)w]
\end{equation}
so the distribution is normalized,
satisfies equation (\ref{eq:pa1}) 
and $w$ is the variance of the distribution
($\langle \alpha^2 \rangle = 1+w$).

The function (\ref{eq:pa_model}) has the attractive
property that when the parameter $w$ spans the interval $0 \le w < \infty$
the quantity $\langle G^2 \rangle$
spans the entire interval of the theoretically allowed values.
For $w=0$ one has 
$\langle G^2 \rangle =\Gamma^2$, and the value
of 
$\langle G^2 \rangle$ 
grows monotonically with $w$, reaching (for $w \to \infty$) the asymptotic
value $\langle G^2 \rangle \to \Gamma$.

The qualitative features of the distributions change with $w$.
\begin{itemize}
\item In the limit of $w\to 0$ the distributions $p(\alpha)$ 
and $p_G(G)$ become delta functions:
$p(\alpha) = \delta [\alpha-1]$ and 
$p_G (G) = \delta [G -\Gamma]$

\item For $w$ small, the distribution $p(\alpha)$
is approximately a gaussian of width $\sigma = \sqrt{w}$ centered
at $\alpha \simeq 1$ while $p_{G}(G)$ is a narrow distribution
centered on $G \simeq \Gamma$.

\item For $0 < w <1$ the distributions $p(\alpha)$ and
$p_G(G)$ have one single maximum. For $p(\alpha)$ the
maximum is at $\alpha = 1-w$, the positions of the maximum 
for $p_G (G)$ depends on $\Gamma$ and $w$.

\item 
For $w=1$ the $p(\alpha)$ distribution is equal to 
$e^{-\alpha}$ while $p_G (G)$ depends monotonically on
$G$ with maximum at $G=0$ ($G=1$) for
$\Gamma < 1/2$ 
($\Gamma >1/2$) (for $\Gamma=1/2$ the distribution is flat). 

\item 
For $w>1$ the $p(\alpha)$ distribution diverges when $\alpha \to 0$,
and decreases monotonically. 
The corresponding $p_G (G)$ has divergence for both 
$G \to 0$ and $G\to 1$ and a single minimum in the center.
When $w \gg 1$ the probability is concentrated in two
small intervals close to $G \simeq 0$ and $G \simeq 1$, and is 
always very small in the remaining central part of the
interval $[0,1]$.

\item
In the limit $w \to \infty$ the 
function $p_G (G)$ takes the asymptotic form:
\begin{equation}
p_G (G) = (1-\Gamma) ~\delta (G) + \Gamma ~\delta(1-G)
\end{equation}
that corresponds to complete transparency or complete absorption.
\end{itemize}

The general form of $p_G (G)$ for an arbitrary value 
of $w$ is:
\begin{equation}
p_G (G) = \frac{[(1-\Gamma)^{-w}-1]^{-1/w}}{\Gamma_{\rm E}(1/w)} 
\left [-\ln (1 -G) \right ]^{-1 + 1/w}
~\left [1 -G \right ]^{-1 + 1/((1-\Gamma)^{-w}-1)} \;,
\end{equation}
for large $w$ this becomes approximately
\begin{equation}
p_G (G) \propto 
\left [-\ln (1 -G) \right ]^{-1} ~\left [1 -G \right ]^{-1} 
\end{equation}
with divergences at $G=0$ and $G=1$.

\clearpage

\end{document}